\documentclass{article}
\usepackage{graphicx} 
\usepackage[affil-it]{authblk}
\usepackage[utf8]{inputenc}
\usepackage[english]{babel}
\usepackage{titlesec}
\usepackage[margin=0.5in]{geometry}
\usepackage{mathtools}
\usepackage{enumerate}
\usepackage{enumitem}
\usepackage{lineno} 
\usepackage{adjustbox}
\usepackage{multirow}
\usepackage{chngcntr}
\usepackage{titlesec}
\usepackage[utf8]{inputenc}
\usepackage{mathrsfs}
\usepackage{lscape}
\usepackage{cite}

\usepackage{cite}

\usepackage{enumitem}
\usepackage{subfigure}
\usepackage{subcaption}

\usepackage[hyphens]{url}
\usepackage{hyperref}
\usepackage{float}
\usepackage[section]{placeins}
\usepackage[dvipsnames]{xcolor}
\usepackage{soul}
\usepackage{amsmath}
\usepackage[labelfont=bf,font=small]{caption}
\usepackage{bm}
\usepackage[title]{appendix}
\usepackage{graphicx}
\usepackage{tabularx}
\usepackage{float}
\usepackage{mathtools}
\usepackage{amsmath,amssymb,amsthm}
\usepackage{oubraces}
\numberwithin{equation}{section}
\usepackage{chngcntr}
\usepackage{mathrsfs}
\usepackage{multirow}
\usepackage{chngcntr}
\usepackage{parskip}
\usepackage{accents}
\usepackage{soul}
\usepackage{orcidlink}

\usepackage{caption}

\usepackage[dvipsnames]{xcolor}   

\definecolor{dgn}{rgb}{0.0, 0.5, 0.0}

\newcommand{\allblack}{\color{black}{}}

\newcommand{\irem}{$|\mathrm{IREM}|$}
\newcommand{\rempre}{$|\mathrm{REMpre}|$}
\newcommand{\rempost}{$|\mathrm{REMpost}|$}

\usepackage{lipsum}
\makeatletter
\g@addto@macro{\endabstract}{\@setabstract}
\newcommand{\authorfootnotes}{\renewcommand\thefootnote{\@fnsymbol\c@footnote}}%
\makeatother

\begin{document}
\title{A Data-Driven Measure of REM Sleep Propensity for Human and Rodent Sleep}
\author{Naghmeh Akhavan \orcidlink{0000-0002-9474-4486}$^{a}$, Alexander G. Ginsberg$^{b}$, Madelyn E. C. Cruz$^{a}$, Yunxi Yan$^{a}$, Shelby R. Stowe\orcidlink{0000-0003-2897-4248}$^c$, Dinesh Pal$^{d}$, Franz Weber$^{e}$, Cecilia G. Diniz Behn \orcidlink{0000-0002-8078-5105}$^{c,f*}$,  and Victoria Booth \orcidlink{0000-0003-2586-8001}$^{a,d}\footnote{Corresponding authors:  {\it vbooth@umich.edu, cdinizbe@mines.edu}}$\\
{\it {\small $^{a}$  Department of Mathematics, University of Michigan, Ann Arbor, MI, USA}}

 {\it {\small $^{b}$  Department of Mathematics, University of Utah, Salt Lake City, UT, USA}}

 {\it {\small $^{c}$  Department of Applied Mathematics and Statistics, Colorado School of Mines, Golden, CO, USA}}
 
 {\it {\small $^{d}$  Department of Anesthesiology, University of Michigan, Ann Arbor, MI, USA}}

 {\it {\small $^{e}$  Department of Neuroscience, University of Pennsylvania, Philadelphia, PA, USA}}

{\it {\small $^{f}$  Department of Pediatrics, University of Colorado Anschutz Medical Campus, Aurora, CO, USA}}

\vspace{-1.5cm}
    }
  \date{}
\maketitle
\vspace{1cm}
\begin{center}
    \textbf{Abstract}
\end{center}
Mammalian sleep is characterized by multiple alternations between episodes of rapid-eye-movement sleep (REMS)  and non-REM sleep (NREMS). While the mechanisms governing the timing of these ultradian NREMS-REMS cycles remain poorly understood, the phenomenon of REMS pressure, namely a drive for REMS that builds up between REMS episodes, is thought to be a contributing factor. {Prior analyses of NREMS-REMS cycles in mice has suggested that time in NREMS is a primary contributor to REMS pressure.} Building on that finding, we previously introduced a REMS propensity measure defined as the probability to enter REMS before the accumulation of an additional amount of NREMS. Analyzing mouse ultradian cycle data, we showed that REMS propensity at REMS onset was positively correlated with REMS bout duration and with the probability of the occurrence of a  REMS bout followed by a short inter-REMS interval,  called a sequential REMS cycle. In this paper, we extend {our} analyses of REMS propensity to human and rat ultradian NREMS-REMS cycle data. {We show that, as in mice, human and rat sleep contain both short NREMS-REMS sequential cycles and longer single NREMS-REMS cycles, though there are some differences in the relative distributions of cycle durations. Although rodents exhibit polyphasic sleep in contrast with the consolidated sleep of humans, the calculated REMS propensity measures in all three species show similar profiles as functions of time spent in NREMS:} specifically, REMS propensity increases with time spent in NREMS  until it reaches a peak value, and then it decays with additional time in NREMS. Positive correlations of REMS propensity at REMS onset with REMS bout duration were present in both human and rat data as in mouse data, suggesting that time spent in NREMS also influences REMS duration in these species. 
Since REMS in humans is known to be influenced by the circadian rhythm, we also investigated circadian modulation of the expression of sequential and single NREMS-REMS cycles in the human data. Specifically, we analyzed REMS expression across the sleep episode and characterized the variation in the occurrence of sequential and single cycles throughout the sleep episode, revealing that nuanced changes in REMS micro-architecture contribute to an increase in percent time spent in REMS as the sleep episode progresses. Overall, our results suggest that similarities in the regulation of NREMS-REMS alternation exist, despite temporal differences, in nocturnal polyphasic rodent sleep and diurnal monophasic human sleep.

\noindent





\section{Introduction}

In mammals, sleep 
alternates between  episodes of rapid-eye-movement sleep (REMS) and non-REM sleep (NREMS), often referred to as ultradian cycling. Although much progress has been made in identifying the hypothalamic and brainstem areas and circuits that are involved in promoting or suppressing REMS \cite{LUPPI2024101907}, the mechanisms governing the timing of ultradian alternation between NREMS and REMS remain poorly understood \cite{le2021asymmetrical}. 

One feature of NREMS-REMS alternation is a bimodal distribution of inter-REMS intervals, defined as the durations of intervals between  successive REMS bouts \cite{zamboni1999control, park2021probabilistic, ursin1970sleep,  kripke1968nocturnal, gregory2002two, amici1994pattern}. In rats and mice, the bimodal distribution  of inter-REMS intervals has led to the identification of two distinct types of NREMS-REMS cycles: {\it single REMS cycles},  defined as REMS bouts that are preceded and followed by longer  inter-REMS intervals, and {\it sequential REMS cycles}, defined as REMS bouts that are separated by shorter inter-REMS intervals and can occur consecutively in  sequences \cite{zamboni1999control,park2021probabilistic}. In human sleep, a similar bimodal distribution of inter-REMS intervals has been identified \cite{merica1991study,kobayashi1985sleep, esposito2003,esposito2004}. {However, in human sleep, it has been a common sleep scoring practice to combine series of consecutive REMS bouts that occur in quick succession (typically separated by less than 15 minutes of NREMS, wake, or movement) into one, consolidated REMS bout \cite{merica1991study}. This scoring practice obscures fragmentation  within ``consolidated'' REMS episodes that is typical in healthy human sleep. However, it remains unclear if these periods of fragmented REMS are comparable to the sequential REMS cycles observed in rodent sleep.}

An active hypothesis for a mechanism governing the timing of NREMS-REMS alternation posits that a short-term homeostatic drive or REMS pressure contributes to the initiation of REMS \cite{beningtonheller1994,beningtonheller1994review,vivaldi1994short,zamboni1999control}. Specifically, a drive for REMS builds up between REMS bouts and discharges during REMS, 
often referred to as “hourglass-like” dynamics \cite{zamboni1999control, park2021probabilistic, heller2021regulation}. This theory is supported by wide-spread evidence that longer REMS bouts are followed by longer intervals before the next REMS bout occurs \cite{barbato1998homeostatic, vivaldi1994short, benington1994remdep,  vivaldi2005short, park2021probabilistic, cajochen2024ultradian}, presumably because more of the drive for REMS is discharged during a longer REMS episode. However, the biological substrate mediating this growth and decay of REMS pressure is unknown.


Various measures that may reflect REMS pressure have been proposed \cite{benington1994remdep, benington1995apamin, bassi2009time, nielsen2010rem, chang2015evening}. 
In an analysis of spontaneous sleep in mice, Park et al.~\cite{park2021probabilistic} proposed a data-driven, probabilistic measure of propensity for REMS, namely the cumulative distribution function (CDF) of the amount of NREMS between REMS bouts.
They found that the CDF of the amount of NREMS between REMS bouts was predictive for the duration of REMS episodes~\cite{park2021probabilistic}. However, in a quantitative sense, REMS pressure should dictate the propensity or probability of entering REMS at a specific time during a sleep episode. Therefore, our group recently proposed an alternative REMS propensity measure, $P(t, \Delta),$ defined as the probability that, after the accumulation of $t$ s of NREMS since the last REMS bout, a transition to REMS will occur within the next $\Delta$ s spent in NREMS \cite{ginsberg2024predictive}. Thus, $P(t, \Delta)$ was the first predictive measure of REMS pressure that is readily interpretable as the probability of entering REMS at a certain time during a sleep episode.
Computing our REMS propensity measure from the same spontaneous sleep data in mice as was used in the Park et al.\ study \cite{park2021probabilistic}, we showed that, as the amount of time spent in NREMS increases, this REMS propensity measure increases until it reaches a peak value. After this point, REMS propensity eventually decays to zero as the time spent in NREMS continues to accumulate. We found that during the light phase, this REMS propensity measured at REMS onset was positively correlated with the duration of the REMS bout \cite{ginsberg2024predictive}. Further, higher propensities following single REMS cycles \cite{zamboni1999control, park2021probabilistic} were correlated with a higher probability of being followed by a sequential REMS cycle \cite{zamboni1999control, park2021probabilistic, ursin1970sleep, merica1991study, kripke1968nocturnal, gregory2002two, amici1994pattern}. However, after the propensity reaches its peak value, its correlation with the 
features of the subsequent REMS bout was lost, suggesting that the amount of time in NREMS drives transitions into REMS only for a limited range of NREMS accumulation.

\begin{figure}[ht!]
    \centering
    \includegraphics[width=1\linewidth]{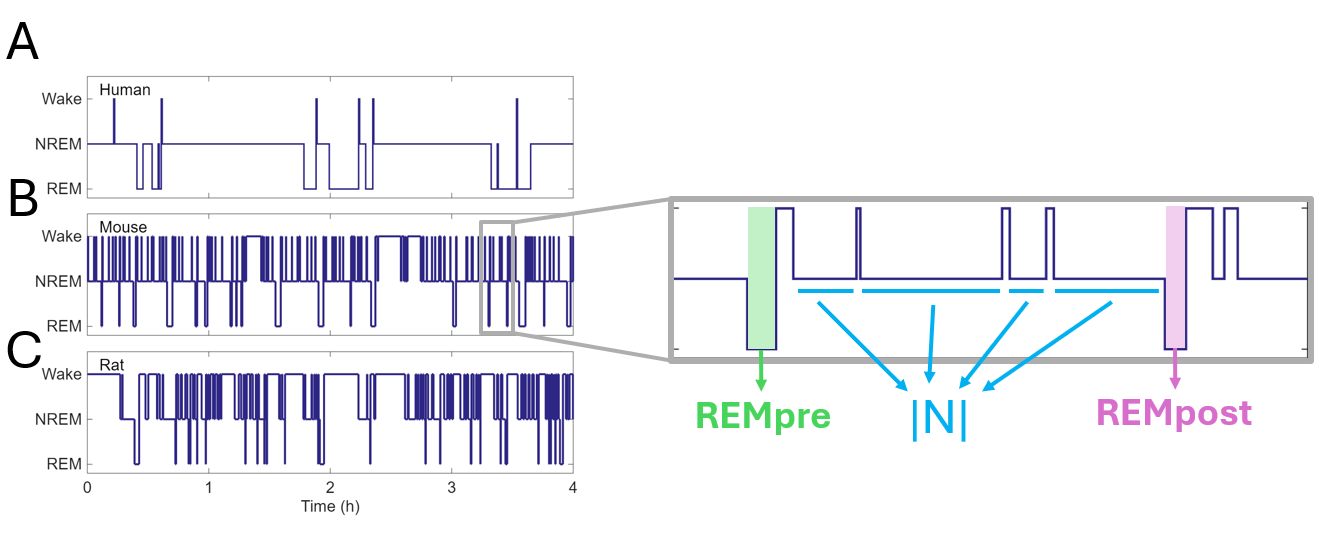}
    \caption{Example hypnograms of sleep behavior over 4 h in human (A), mouse (B, light period) and rat (C, light period). The inset highlights a representative REMS cycle that is initiated at the onset of a REMS bout, REMpre (green), and terminates at the onset of the next REMS bout, REMpost (magenta). The total time spent in NREMS during the inter-REMS interval is defined as $|N|$ (blue). {Note that the full inter-REMS interval contains both NREMS and wake bouts shorter than 2 minutes.} }
    \label{fig:hypno_species}
\end{figure}

{In this paper, we extend our analysis of REMS propensity to (diurnal) human and (nocturnal) rat sleep data and present it in the context of the original mouse data from the Weber lab \cite{park2021probabilistic}. We used a comparable set of rat sleep data recorded in the Pal lab \cite{silverstein_ratdata}. For the human data, we utilized three publicly available human sleep datasets \cite{bitbrain_dataset,sleep_edf_goldberger,mnc_zhang} comprised of five study cohorts. These datasets were collected in participants representing a range of demographic groups, over different time periods, under different inclusion/exclusion criteria, and with different experimental and scoring protocols (for details, see Methods and \cite{bitbrain_dataset,sleep_edf_goldberger,mnc_zhang, sleep_edf_kemp, stephansen2018neural}). 

For our analysis, }we first compare characteristics of NREMS-REMS alternation across human, mouse and rat species. Then we show that the bimodal distribution of time spent in NREMS between REMS bouts in human and rat data can be fit with 
mixture models (MMs). The rat data is fit well with a Gaussian mixture model, similar to  the mouse  data, while the human data is fit with a three part mixture model. Using the MM fits, we compute the REM propensity functions for each species and find that, despite differences in the MMs, it increases with time spent in NREMS up to a peak value and then decreases with further NREMS accumulation. We show that in all three species the duration of a REMS bout is positively correlated with the REMS propensity value at REMS onset, before the propensity reaches its peak value.

The circadian timing of REM sleep varies between nocturnal and diurnal animals with REM sleep most common during the inactive period. This is particularly pronounced in humans who typically experience REMS only during consolidated nocturnal sleep and show circadian effects within the sleep episode. To establish circadian features of REMS expression in the human data, 
we analyzed how the occurrence of sequential and single REMS cycles evolves across the sleep episode. We find that sequential cycles occur more often at the very beginning and at the very end of the sleep episode, revealing nuanced changes in REMS micro-architecture that contribute to an increase in percent time spent in REMS as the sleep episode progresses. 

Overall, our results highlight similarities in NREMS-REMS alternation across three mammalian species, human, rat, and mouse, regardless of the temporal differences between nocturnal polyphasic rodent sleep and diurnal monophasic human sleep.  Furthermore, our work shows how these similarities may be masked when sequential REMS cycles are combined following standard practices for scoring human sleep.

\section{Results}
To investigate the conservation of a probabilistic structure of NREMS-REMS cycling across mammalian species, we analyzed sleep-scored data from human, mouse and rat sleep recordings using a unified computational framework. 
For each species, we segmented sleep into successive REMS cycles, each comprised of an initial REMS bout, REMpre, followed by an inter-REMS interval containing NREMS and wake states (human data may also contain epochs scored as movement) (Figure \ref{fig:hypno_species}). 
{To distinguish overall cycle timing from the amount of NREMS that we posit is specifically relevant to REMS pressure, we consider two related durations within each REMS cycle. 
We define the duration of the inter-REMS interval, denoted by \irem, as the total time from the end of one REMS bout to the onset of the next REMS bout. 
This interval includes all intervening states, including NREMS and any brief wake or movement periods that do not meet the long-wake exclusion criterion (see below). 
Within this interval, we define $|N|$ as the cumulative time spent in NREMS only. 
Thus $|N| \le$  \irem, with equality only when no wake or movement occurs in the interval. 
We consider both quantities because they address different biological questions: \irem \ characterizes the overall spacing of REMS bouts and the temporal structure of ultradian cycling, whereas $|N|$ isolates the NREMS accumulation hypothesized to contribute to REMS pressure. 
Since our propensity measure is intended to quantify the probability of entering REMS as a function of prior NREMS accumulation, $|N|$ is the quantity used in our analysis. }


Before performing any statistical or model-based analysis, all datasets were preprocessed using a long-wake (LW) filtering criterion. Because extended spontaneous wake episodes can disrupt the intrinsic timing of NREMS-REMS alternation, we excluded any REMS cycle whose inter-REMS interval contained a contiguous wake segment of at least 2 minutes. This threshold was applied identically to human data and rodent data (in both light and dark conditions), ensuring consistent cross-species comparisons. The choice of the 2-minute cutoff was based on a systematic evaluation of alternative thresholds ($2,5,7,10$ minutes) which demonstrated that the LW threshold set to 2 minutes resulted in the most stable and well-structured distribution of NREMS durations for each species; full justification and sensitivity analyses are provided in Supplementary Section~\ref{sec: supp_filter}. 
{In addition, due to the heterogeneity of participants, protocols, and sleep scoring in the human data, we applied a subject-level outlier filter based on the number of valid REMS cycles exhibited per subject after LW filtering for the human data only. Sleep records with cycle counts outside the interquartile-range criterion ($< Q_1-1.5\,\mathrm{IQR}$ or $> Q_3+1.5\,\mathrm{IQR}$) were excluded to reduce disproportionate influence from outliers (see Methods section for details)}.

\subsection{Basic Statistics of NREMS–REMS Alternation Across Species}

We first summarize metrics of NREMS–REMS cycling for each species. Considering metrics averaged across individuals (Table~\ref{tab:rem cycle stat}), the average number of REMS cycles per hour varied across species and light-dark condition, with humans, mice, and rats all exhibiting 
on the order of 2--6 cycles per hour when averaged per recording.
Humans showed $5.04\pm 5.44$ REMS  episodes per hour, reflecting substantial inter-individual variability. The similar values of the mean and standard deviation reflect a high variability in the average REMS episode rate when fragmented REMS bouts are not consolidated. 
{These cross-species differences in variability should be interpreted with caution. The human dataset included substantially more analyzed recordings than the rodent datasets and was assembled from multiple public cohorts with differing acquisition and scoring protocols. Accordingly, the larger standard deviations observed in the human data likely reflect both genuine heterogeneity in human REMS organization and broader between-recording variability arising from the larger and more heterogeneous dataset, rather than a simple species difference alone.}
Mice displayed clear light--dark modulation, with higher cycling rates during the light phase ($5.04\pm 1.74$ cycles/hour) and slower cycling during the dark phase ($2.24\pm 0.60$ cycles/hour).
Rats exhibited comparable cycling rates in both phases, averaging $5.77\pm 1.71$ cycles per hour in the light phase and $5.67\pm 1.62$ cycles per hour in the dark phase. 
Mean REMS cycle duration ranged from approximately $10$--$30$ minutes in rodents to roughly 25--30 minutes in humans. For rodents, REMS occupied smaller fractions of the cycle duration with NREMS and wake approximately splitting the remainder of the cycle length in the light period. Since wake episodes within a cycle are limited by the 2 minute LW threshold, this reflects many brief wake episodes interrupting sleep in rodents. In humans, REMS occupied a larger fraction of the cycle and the inter-REM interval consisted primarily of NREMS.

\begin{table}[H]
    \centering
    \begin{tabular}{|c|c|c|c|c|c|c|}
        \hline
        & & \multicolumn{4}{|c|}{Means of Individual Recording Means} & \\
        \hline
        \multirow{2}{*}{Species} & Number of & REMS episodes  & \multirow{2}{*}{\rempre} & \multirow{2}{*}{$|N|$}  & \multirow{2}{*}{\irem}  & REMS cycle  \\ 
                                 & recordings & per hour    &  &      &  & duration   \\
        \hline
        Human 
        & 515 
        & $5.04 \pm 5.44$ 
        & $10.41 \pm 7.68$ 
        & $16.30 \pm 20.48$ 
        & $16.80 \pm 20.55$ 
        & $27.21 \pm 25.85$ \\ 
        \hline

        Mouse (Light) 
        & 179 
        & $5.04 \pm 1.74$ 
        & $1.02 \pm 0.23$ 
        & $7.83 \pm 2.49$  
        & $12.43 \pm 4.87$ 
        & $13.46 \pm 5.03$ \\ 
        \hline

        Mouse (Dark) 
        & 54 
        & $2.24 \pm 0.60$ 
        & $1.10 \pm 0.19$ 
        & $10.39 \pm 2.92$  
        & $27.67 \pm 7.77$ 
        & $28.77 \pm 7.85$ \\ 
        \hline

        Rat (Light) 
        & 44 
        & $5.77 \pm 1.71$ 
        & $1.03 \pm 0.37$ 
        & $4.44 \pm 1.25$  
        & $10.51 \pm 4.05$ 
        & $11.55 \pm 4.17$ \\ 
        \hline

        Rat (Dark) 
        & 37 
        & $5.67 \pm 1.62$ 
        & $1.05 \pm 0.36$ 
        & $4.43 \pm 1.32$  
        & $10.66 \pm 4.16$ 
        & $11.71 \pm 4.25$ \\ 
        \hline
    \end{tabular}

    \caption{{\bf REMS cycle summary statistics (in minutes) averaged per recording for each species.}
    Values are mean $\pm$ standard deviation (SD). ``REMS  episodes per hour'' is computed as (cycle count)/(analyzed time).
    \rempre \ = duration of the REMS bout initiating the cycle.
    $|N|$ = cumulative time in NREMS during the inter-REMS interval (excluding wake and movement; sequential cycles may have $|N|=0$).
    \irem \ = total time from the end of the REMS episode initiating the cycle to the beginning of the next REMS episode, including NREMS, wake and all other epochs such as movement. 
    REMS cycle duration = \rempre \ + \irem.
    All datasets were filtered with a long-wake episode threshold of 2 minutes within the inter-REMS interval.}
    \label{tab:rem cycle stat}
\end{table}


To assess REMS cycle features at the population level, 
we also examined statistics when all cycles for a species were pooled together (Table~\ref{tab:rem pooled cycle stat}). 
Similar cross-species scaling reappears: per cycle, humans displayed the longest inter-REMS intervals ($24.23 \pm 35.42$ minutes), compared with $5.88 \pm 4.08$ minutes in mice (light phase) and $3.30 \pm 3.58$ minutes in rats (light phase). 
As also apparent in the individual averages, inter-REMS interval duration (\irem ) and the NREMS-only component ($|N|$) of the inter-REMS interval were especially variable in humans where the SD may exceed the mean, indicating substantial heterogeneity on REM cycle durations. 
Rodents showed shorter timescales but comparable relative variability in \irem \ and $|N|$: coefficients of variation were approximately $0.7$ in mice and $0.9$–$1.1$ in rats, indicating that inter-REMS durations remain heterogeneous even within shorter ultradian cycles.

\begin{table}[H]
    \centering
    \begin{tabular}{|c|c|c|c|c|c|}
        \hline
        & & \multicolumn{4}{|c|}{Pooled REMS cycles for each species} \\
        \hline
        \multirow{2}{*}{Species} & Total REMS  & \multirow{2}{*}{\rempre}  & \multirow{2}{*}{$|N|$}  & \multirow{2}{*}{\irem}  & REMS cycle  \\ 
                                 & cycles     &  &                        &    & duration   \\
        \hline
        Human         & 2936/4426 & $8.89 \pm 9.41$  & $22.68 \pm 33.20$ & $24.23 \pm 35.42$ & $33.12 \pm 39.89$  \\ \hline
        Mouse (Light) & 4005/5300 & $0.84 \pm 0.74$  & $5.16 \pm 3.72$   & $5.88 \pm 4.08$   & $6.72 \pm 4.57$    \\ \hline
        Mouse (Dark)  & 739/1307  & $0.91 \pm 0.68$  & $5.94 \pm 3.76$   & $6.82 \pm 4.10$   & $7.72 \pm 4.53$    \\ \hline
        Rat (Light)   & 2089/2766 & $0.88 \pm 1.03$  & $2.82 \pm 3.25$   & $3.30 \pm 3.58$   & $4.17 \pm 4.07$    \\ \hline
        Rat (Dark)    & 1745/2312 & $0.89 \pm 1.04$  & $2.87 \pm 3.27$   & $3.36 \pm 3.60$   & $4.24 \pm 4.09$    \\ \hline
    \end{tabular}
    \caption{
    {\bf REMS cycle summary statistics (in minutes) averaged per cycle for each species.}
    For each species and light condition, means and SDs are computed for REMS cycles in all recordings combined together.
    Total REMS cycles reports (number of REMS cycles after the \(\ge 2\)-minute LW filtering threshold was applied)/(total number of REMS cycles).}
    \label{tab:rem pooled cycle stat}
\end{table}

{The differences between Table~\ref{tab:rem cycle stat} and Table~\ref{tab:rem pooled cycle stat} arise from the different weighting schemes used in the two summaries. 
Table~\ref{tab:rem cycle stat} averages recording-level means and therefore gives equal weight to each recording, whereas Table~\ref{tab:rem pooled cycle stat} averages across all pooled REMS cycles and therefore gives equal weight to each cycle. 
As a result, recordings containing larger numbers of cycles contribute more strongly to the pooled averages in Table~\ref{tab:rem pooled cycle stat}. 
In rodents, this leads to substantially shorter pooled \irem \ values, indicating that recordings with more frequent REMS cycling tend to have shorter inter-REMS intervals and thus dominate the cycle-level averages. 
In humans, the opposite pattern is observed, with pooled \irem \ slightly exceeding the mean of recording-level means, suggesting a different relationship between number of cycles and inter-REMS timing across recordings. 
These differences highlight substantial heterogeneity across recordings and show that recording-level and cycle-level summaries capture distinct aspects of REMS organization.}

\subsection{Relationship between preceding REMS duration and subsequent inter-REMS interval}
\begin{figure}
    \centering
    \includegraphics[width=1\linewidth]{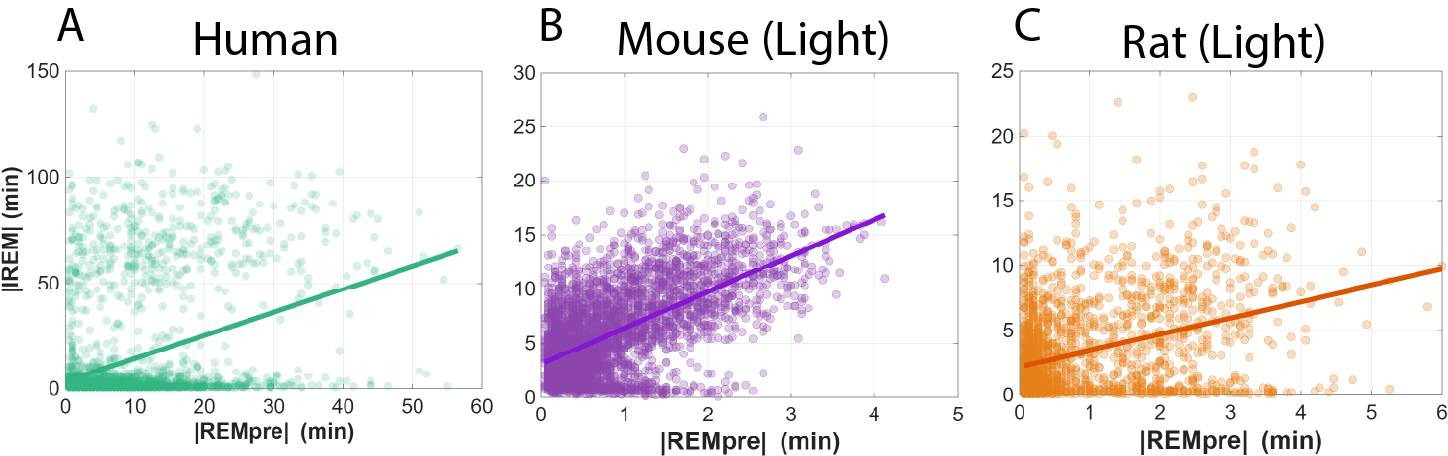}
    \caption{{\bf Inter-REMS interval duration increases with the duration of the preceding REMS bout across species.}  Positive correlations are exhibited between REMS bout duration \rempre\ and the subsequent inter-REMS interval duration \irem \ in all species: (a) human (green), (b) mouse (light phase, purple), and (c) rat (light phase, orange). Each point represents one REMS cycle; solid  lines are least-squares fits  with slopes $1.085, 3.339$, and $1.262$ for human, mouse (light), and rat (light), respectively.
    The reported significance levels were obtained from a two-sided test of zero Pearson correlation using the Fisher $z$-transform normal approximation, with null hypothesis $H_0: \rho=0$ and alternative hypothesis $H_A:\rho \neq 0$, where $\rho$ denotes the population correlation between \rempre\ and \irem.
    The corresponding $p$-values are $p = 6 \times 10^{-94}$ (human), $p \ll 0.001$ (mouse), and $p=1.68 \times 10^{-67}$ (rat).
    The mouse and rat dark phases are shown in Supplementary Figure~\ref{fig: S1_dark}.}
  \label{fig:scatter_plots}
\end{figure}

We next evaluate whether the duration of a REMS bout influences the length of the subsequent inter-REMS interval as has been previously reported \cite{le2021asymmetrical,vivaldi2005short} and as predicted for a REMS pressure process with hourglass-type dynamics. In Figure~\ref{fig:scatter_plots}, we compare $|\mathrm{IREM}|$ with $|\mathrm{REMpre}|$ in all REMS cycles for each species (see supplementary Figure~\ref{fig: S1_dark} for dark phase rodent data). 
Each point in the scatter plots represents a single REMS cycle, with the x-axis denoting the duration of the REMS bout initiating the cycle, \rempre, and the y-axis denoting the subsequent inter-REMS duration \irem. Linear least-squares fits (solid lines) quantify the direction and strength of association between these variables. 

Across all species, longer REMS bouts were generally followed by longer inter-REMS intervals, producing positive correlations consistent with an hourglass-type homeostatic process in which a longer REMS episode discharges a greater portion of accumulated REMS pressure, thereby delaying the onset of the next REMS episode. This trend was most pronounced in mice (Figure~\ref{fig:scatter_plots}b), where the regression slope indicates that each additional minute of REMS was followed on average by several minutes of additional 
inter-REMS time. 
While the trend line slopes for human and rat data were positive, the data showed higher variability for this relationship.
Together, the results confirm that the duration of a preceding REMS episode predicts the length of the subsequent inter-REMS interval across species, supporting the hypothesis that REMS is regulated by a  homeostatic process that resets during each bout. 

\subsection{Modeling inter-REMS interval distributions}\label{sec: 2.3}
\begin{figure}
    \centering
    \includegraphics[width=1\linewidth]{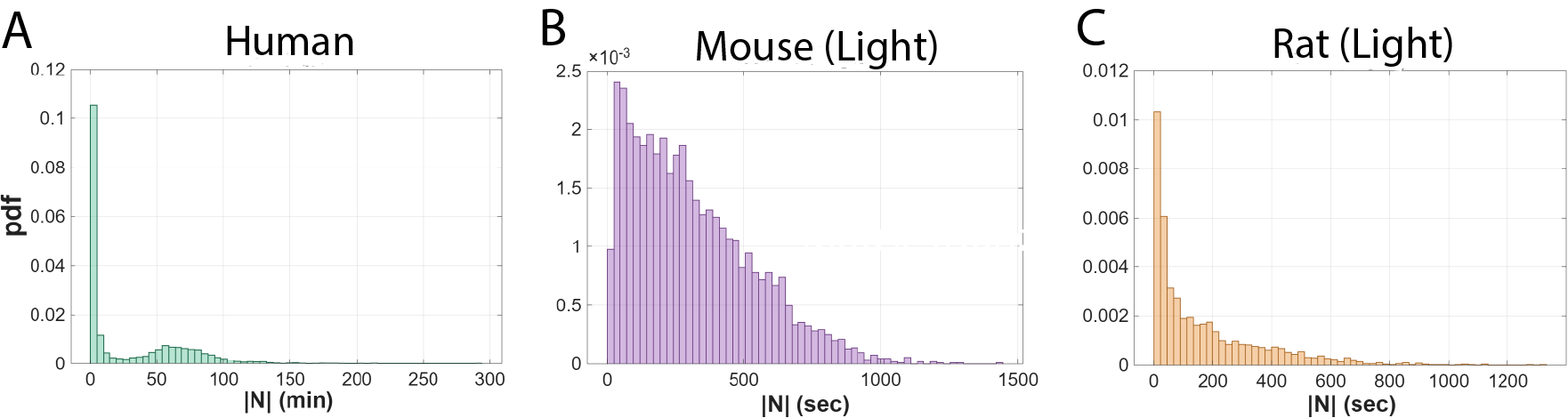}
 \caption{{\bf Empirical inter-REMS $|N|$ distributions across species suggest multiple characteristic timescales. } 
 Histograms show the empirical distribution of $|N|$, where $|N|$ is the cumulative duration of NREMS in the inter--REMS interval, for (a) human, (b) mouse (light phase), and (c) rat (light phase) data (see Supplementary Figure~\ref{fig: S2_dark} for dark phases of rodent data). 
 Across species, $|N|$ spans a wide range and exhibits structured, non-Gaussian profiles. 
 }
  \label{fig:gmm_lumped_plots}
\end{figure}

\begin{figure}[ht!]
    \centering
    \includegraphics[width=1\linewidth]{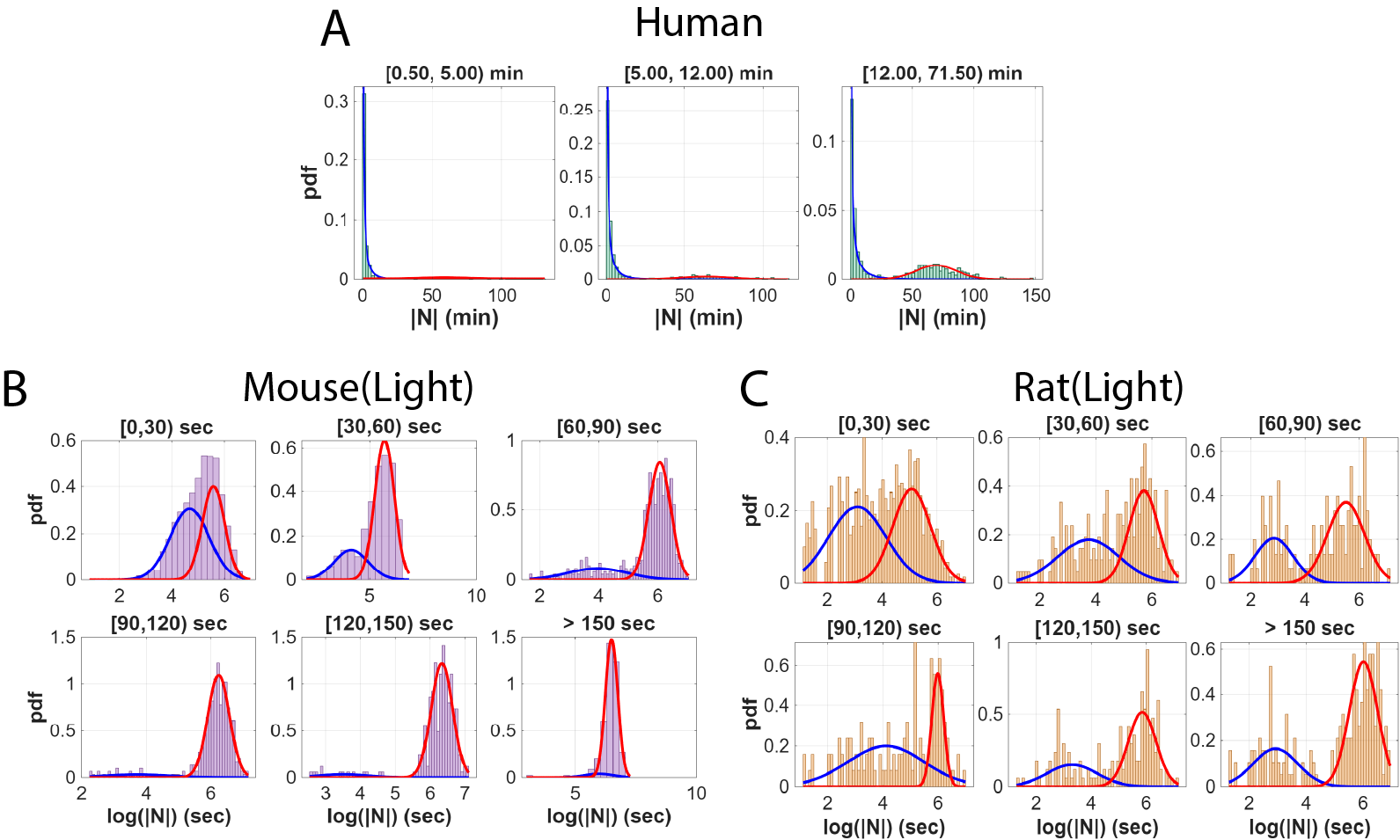}
    \caption{\textbf{Mixture model fits of inter-REMS $|N|$ distributions partitioned by \rempre \ duration.}
Mixture models fitted to $|N|$ distributions for human data (a) and $\log(|N|)$ distributions for rodent data (b: mouse (light phase), c: rat (light phase)) for REMS cycles with similar durations of the preceding REMS bout \rempre; dark phase results for mouse and rat data are provided in Supplementary Figure~\ref{fig: S3_dark}.
Each subpanel corresponds to a distinct \rempre \ bin (ranges in seconds or minutes shown in panel titles).
Histograms show empirical pdfs, where blue and red curves denote the short- and long-interval components of the mixture model. }
  \label{fig:gmm_bin}
\end{figure}

To characterize the statistical structure of inter-REMS durations across species, we examined the distribution of cumulative NREMS sleep in inter-REMS intervals, quantified by $|N|$ {(Figure~\ref{fig:gmm_lumped_plots})}. 
Despite differences in absolute timing and distributional shape across species, all three distributions span both short and long inter--REMS intervals. 
The human and rat data exhibit a pronounced enrichment of short $|N|$ durations with a broad right tail at longer $|N|$. This pattern is consistent with earlier studies showing that inter--REM intervals cluster into two dominant time scales, corresponding to short ``sequential” and longer ``single” REMS cycles \cite{esposito2004, barbato1998homeostatic, zamboni1999control}. When all REMS cycles are considered together in the mouse data, this pattern is less obvious, instead exhibiting a broad, skewed distribution across a wide range of $|N|$ durations with considerable overlap between shorter and longer durations.  However, following previous work \cite{park2021probabilistic, ginsberg2024predictive}, we partitioned REMS cycles by \rempre\ and for the rodent data considered the distribution of 
$\log(|N|)$ which revealed bimodal $|N|$ distributions for all species (Figure~\ref{fig:gmm_bin}).


We modeled these empirical distributions of $\log(|N|)$ for the rodent data and $|N|$ for the human data in each \rempre \ bin using mixture models (MMs, Figure~\ref{fig:gmm_bin}). For both rodent data sets, we fit a two-component Gaussian mixture model (GMM) to 
$\log(|N|)$ in fixed-width \rempre\ bins of $30$~(s), namely $[0, 30)$, $[30, 60)$, $[60, 90), [90, 150)$, and $> 150$~(s) (Figure~\ref{fig:gmm_bin}(b, c)) following previous work \cite{park2021probabilistic,ginsberg2024predictive} (see supplementary Figure~\ref{fig: S3_dark} for rodent dark-phase results). Parameters for the fit GMM functions are listed in Supplementary Tables \ref{tab:gmm_mouse_light}--\ref{tab:gmm_rat_dark}. 
For the human data, REMS cycles were partitioned based on \rempre\ into three bins where \rempre $=[0.5, 5), [5, 12)$, and $[12, 71.5)$ minutes (Figure~\ref{fig:gmm_bin}(a)), and model fits were performed on the three resulting empirical $|N|$ cumulative distributions (in minutes).
Because the human $|N|$ distributions show a non-negligible point mass at the measurement floor $x_{\min} = 30$s, corresponding to the sleep scoring epoch, together with a broad heavy right tail, we did not fit a Gaussian mixture model on $\log(|N|)$. Instead, we modeled the human $|N|$ distributions using a three-part mixture consisting of an atom (point mass) at $x_{\min}$, a short-duration continuous component given by the normalized $E1$ form $\propto e^{-rt}/t$ on $[x_{\min}, \infty)$, and a long-duration truncated normal component. Details of the fitting procedure including likelihood estimation, fitting constraints and bootstrap validation are given in the Supplement~\ref{sec: supp_human_fit}. Parameters for the fit three-part MM are given in Supplementary Table \ref{tab:atom_fit_results}.
 
The MM fits provide a quantitative method to categorize sequential and single REMS cycles \cite{park2021probabilistic,ginsberg2024predictive}. The intersection point of the short and long model components (blue and red curves in Figure~\ref{fig:gmm_bin}) determines a threshold value where REMS cycles with $|N|$ less than the intersection point are considered sequential cycles and those with $|N|$ greater than the intersection point are considered single cycles. In all species
across \rempre\ bins, the longer-duration (``single-cycle") model component systematically shifts to larger values with increasing \rempre, indicating that longer preceding REMS bouts are associated with longer subsequent $|N|$ duration. This pattern is consistent with the positive association between \rempre\ and \irem\ (Figure~\ref{fig:scatter_plots}). For rodent data, the shorter-duration (``sequential-cycle") component varies comparatively little across \rempre\ bins, suggesting that the characteristic timescale of short-interval REMS cycling is relatively stable in rodent sleep.

Across species, the  fitted mixture models closely tracked the empirical $|N|$ distributions. 
Agreement was quantified using Kolmogorv--Smirnov (KS) statistics comparing empirical and fitted CDFs. 
For rodent data, KS-best distances for the 2-component GMM fits to 
$\log(|N|)$ were consistently small at the pooled level ($D= 0.0157$--$0.0499$) and remained modest across REMpre bins (typically $D \lesssim 0.06$; see Table~\ref{tab:ks_distances_ksbest_bins_by_species} in the Supplement), with the largest deviations confined to a small number of tail bins (maximum $D=0.0908$ in Mouse (Dark) Bin~6 and $D=0.0884$ in Rat (Dark) Bin~4). 
Consistent with these distances, KS diagnostic $p$-values exceeded $0.05$ for every rodent dataset in all bin-specific fits (Table~\ref{tab:ks_pvalues_bins_by_species} in the Supplement), indicating no bin exhibited an obvious distributional mismatch. 
For human data, KS distances were determined to be insignificant by refit parametric bootstrapping (see Table~\ref{tab:atom_fit_results} and Supplement \ref{supp:humanGOF} for details).

Together, these diagnostics support the conclusion that mixture-based models provide faithful statistical summaries of inter--REMS NREMS accumulation across species 
and within \rempre \ bins, with the largest departures concentrated in a small number of low-sample, tail-duration bins.
From these fitted distributions we compute a continuous function for the cumulative distribution function (CDF) for $|N|$, $F(|N|)$, which serves as the foundation for calculating the REM-propensity function $P(t, \Delta)$ in the next section.

\subsection{REMS propensity \texorpdfstring{$P(t,\Delta)$}{P(t, Δ)} across species}
Building on the MM representations of $|N|$ distributions, we next compute the REMS propensity function $P(t,\Delta)$ that quantifies the probability of entering REMS in the near future as NREMS accumulates. 
After $t$ seconds of accumulation of NREMS since the last REMS bout, 
the probability that a transition to REMS occurs before an additional $\Delta$ seconds of NREMS is given by
\begin{align}
    P(|N|, \Delta) = \frac{F(|N|+\Delta) - F(|N|)}{1-F(|N|)}, 
    \label{eq:propensity}
\end{align}
where $F(|N|)$ is the cumulative distribution function (CDF) of $|N|$ obtained from the species-specific MM fit functions. We fixed $\Delta= 30$ s to represent a short ``near-future" window similar to prior work 
\cite{ginsberg2024predictive}.  By definition, $P(t, \Delta)$ reflects the instantaneous likelihood that an ongoing NREMS episode will terminate in REMS within the next $\Delta = 30$ s.

\begin{figure}[ht!]
    \centering
    \includegraphics[width=1\linewidth]{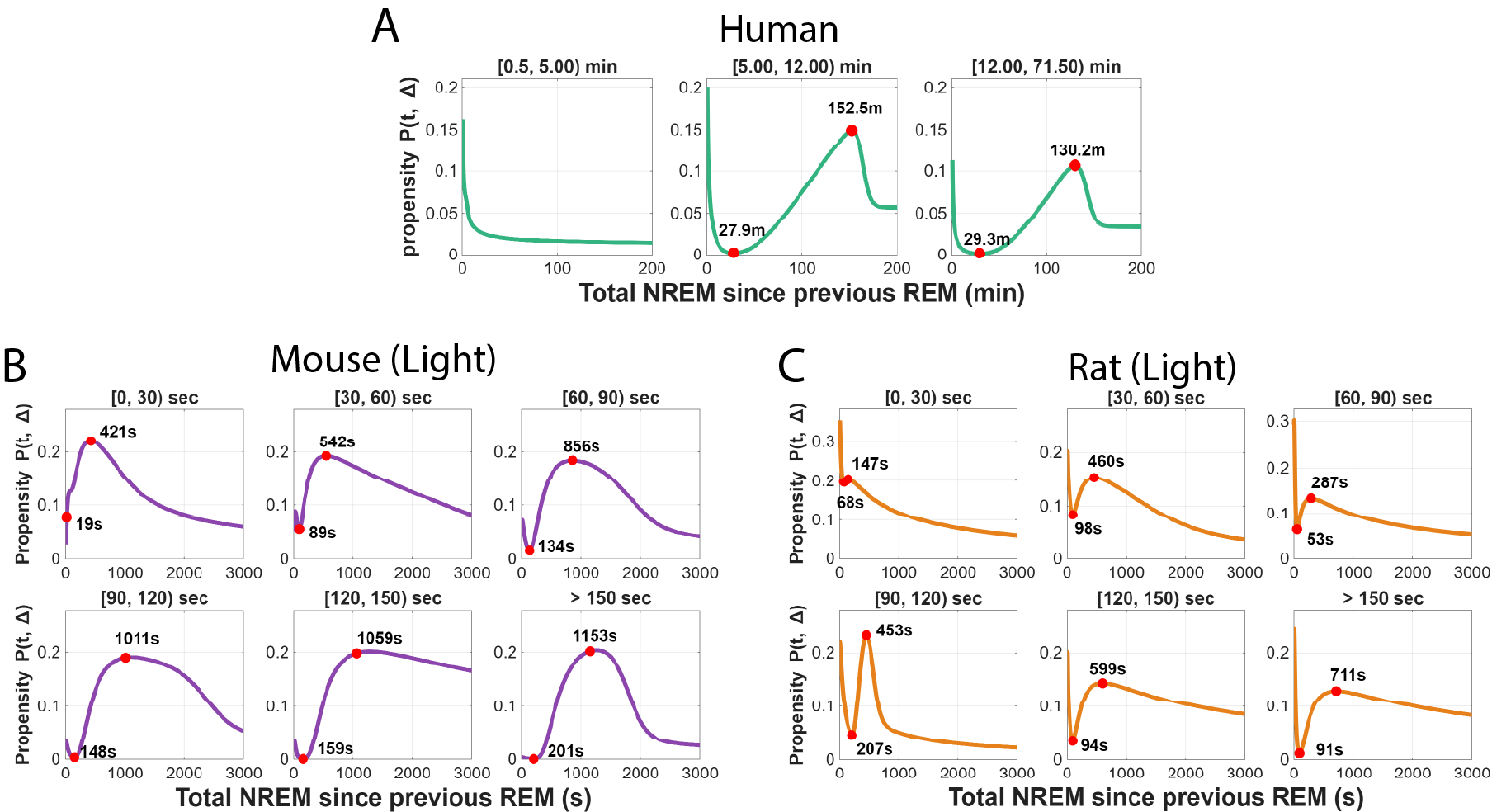}
    \caption{\textbf{REMS propensity functions grouped by \rempre \ durations.}
Propensity functions $P(t,\Delta)$ (Eq.~\eqref{eq:propensity}) {are shown for REMS cycles grouped by similar \rempre\ durations, corresponding to the binned inter--REMS $|N|$ distributions in}
Figure \ref{fig:gmm_bin}, for (a) human, (b) mouse (light phase), and (c) rat (light phase) data. 
{For human data, $P(t, \Delta)$ was computed from the three part atom + E1-short + truncated-normal mixture model fit to $|N|$ (in minutes); for rodents, $P(t,\Delta)$ was computed from the two-component Gaussian mixture model (GMM) fit to $\log(|N|)$ (in seconds).}
In each panel, $P(t,\Delta)$ {is plotted as a function of the accumulated} NREMS duration $t$ since the prior REMS bout. 
Red markers denote local minima (troughs) and maxima (peaks) {with the correponding $|N|$ values annotated. 
Dark-phase mouse and rat results are shown in Supplementary Figure~\ref{fig: S5_dark}}.
}
  \label{fig:propensity_bin}
\end{figure}

Figure~\ref{fig:propensity_bin}  shows the REMS propensity function curves $P(t, \Delta)$ computed from the fitted mixture models, in each  \rempre\ bin, for each species using the inter--REMS $|N|$ distributions in Figure \ref{fig:gmm_bin} 
(mouse and rat dark phase propensity curves are shown in Supplementary Figure~\ref{fig: S5_dark}). 
In all species and in all but the shortest \rempre \ bins, $P(t, \Delta)$ exhibited a characteristic non-monotonic profile: after an initial transient, it increased with NREMS accumulation, reached a clear local maximum, and then decayed with further NREMS accumulation.
This profile suggests that the propensity to enter REMS  increases only up to a finite amount of NREMS accumulation before declining. In other words, once accumulated NREMS time becomes very long, it alone is insufficient to explain REMS timing and additional factors (e.g., circadian phase, arousal/wake intrusions, or other regulatory processes) likely contribute more strongly to the ongoing sleep pattern.

The presence of troughs (red markers) in 
$P(t, \Delta)$ at short values of $|N|$ 
reflects the prevalence of sequential REMS cycles that feature minimal NREMS accumulation in the inter--REMS interval.  
This suggests that $P(t, \Delta)$ may not be applicable to predicting REMS onset in sequential cycles with $|N|$ durations less than its trough value but is more applicable for REM onset prediction in longer single cycles with $|N|$ durations between the trough and the peak.
In the human data, for short \rempre \ durations ($[0.5, 5)$ min), the propensity $P(t, \Delta)$ decreases rapidly with increasing accumulated NREM time and remains low thereafter, with no pronounced interior maximum, consistent with short duration REMS bouts occurring as sequential REMS cycles in which REMS re-entry typically occurs after 
relatively little intervening NREMS. 

{The ranges of $|N|$ (human) or $\log(|N|)$ (rodent) values at the local peaks of $P(t, \Delta)$ (red markers) vary across species. The longest durations occur in humans and the mice show longer durations than the rats,
supporting the interpretation that $P(t, \Delta)$ encodes a species-specific homeostatic timescale for REM pressure.} For the rodent propensity functions, the $|N|$ locations of the peaks shift to larger values with increasing \rempre (with some variability for rats), and these shifts are larger than those observed for trough locations. This pattern mirrors the positive association between \rempre\ and inter--REM interval duration: longer preceding REMS bouts are followed by longer NREMS accumulation before the next REMS episode. This trend was not observed in the propensity functions computed for the two \rempre\ bins considered for the human data.
The occurrence of these peaks in $P(t, \Delta)$ and the shifts in their locations with increasing \rempre\ indicates a characteristic ``preferred" waiting time scale for REMS re-entry following longer REMS bouts, supporting a transition from primarily sequential cycling toward more isolated REMS episodes separated by extended NREMS accumulation. 

Taken together, these results demonstrate that the {temporal organization of REMS onset shares a common} probabilistic structure across mammals, {while the characteristic timescales and slopes 
of this structure are species dependent.}
The non-monotonic shape of $P(t, \Delta)$ is consistent with 
a limited-range {NREMS-dependent drive that increases} 
the likelihood of REMS transitions up to a species-specific duration, after which 
{additional factors beyond accumulated NREMS time likely contribute to REMS timing (see Discussion).}

\subsection{Testing the predictive utility of REMS propensity \texorpdfstring{$P(t,\Delta)$}{P(t, Δ)}}
To assess {whether the } 
REMS propensity measure $P(t, \Delta)$ {carries predictive information about a}
subsequent REMS episode and whether this relationship differs across species, we examined the association between $P(t, \Delta)$ { evaluated} at REMS onset and the duration of the ensuing REMS bout (\rempost).
{We focus on REM cycles whose NREMS accumulation $|N|$ lies in the increasing propensity regime (between the trough and peak of $P(t, \Delta)$; see Methods). In this regime propensity varies systematically with $|N|$ and therefore provides a meaningful continuous predictor.}

\begin{figure}[h!]
    \centering
    \includegraphics[width=1\linewidth]{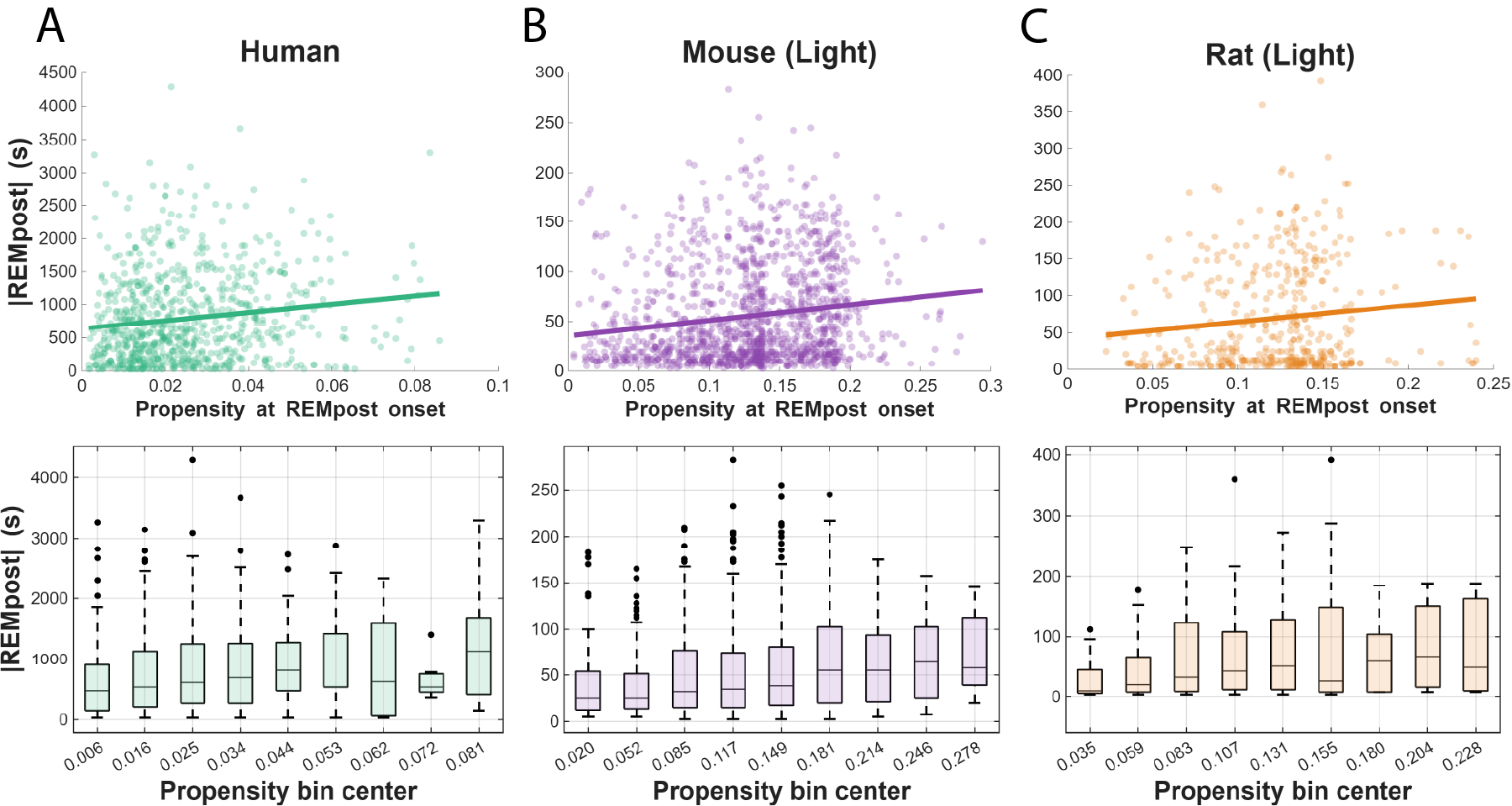}
    \caption{\textbf{
  Correlation between REMS propensity and the duration of the subsequent REMS bout across species.}
    Top row: Scatter plots show the relationship between the REMS propensity at REMS onset and the duration of the following REMS episode ($|\mathrm{REM{post}}|$) (in seconds) for 
    (a) human, (b) mouse (light phase), and (c) rat (light phase) REMS cycles.
    Each point represents a single REMS cycle; and solid lines denote least-squares linear fits computed over cycles within the increasing propensity regime {(human: $p=6.94\times10^{-5}$; mouse light: $p=3.34\times10^{-9}$; rat light: $p=0.0101$)}.
    Bottom row: Box-and-whisker summaries of $|\mathrm{REM}{\mathrm{post}}|$ distributions across binned propensity values for (a) human, (b) mouse (Light), and (c) rat (Light). Boxes denote the interquartile range with a median line; whiskers extend to $1.5\times$IQR beyond each end of the interquartile range; points denote outliers.
    }
  \label{fig:propensity_rempost_scatter}
\end{figure}

{Figure~\ref{fig:propensity_rempost_scatter} summarizes this relationship across species. The top row (a--c) shows cycle-by-cycle scatter plots of \rempost\ versus the propensity value at REMS onset for human, mouse (light phase), and rat (light phase) REMS cycles. In all three datasets, the fitted regression lines slope upward, indicating a consistent positive association: REMS episodes that begin at higher propensity values tend to be followed by longer REMS bouts. The bottom row provides a complementary nonparametric view where REMS cycles are binned according to the propensity value at their onset.   Across species, the box-and-whisker summaries of the corresponding \rempost\ distributions show an overall upward shift in median values (lines across boxes) with increasing propensity bin values. This confirms that the positive trend is not driven by a small number of outliers but reflects a broad shift in the typical (\textit{median}) \rempost\ as propensity increases. 
Analogous analysis for rodent dark phase data is shown in Supplementary Figure~\ref{fig: S6_dark}. }

{Notably, the strength and dynamic range of this relationship differ across species. Human cycles span a much wider range of \rempost\ values (including rare very long bouts), whereas rodents exhibit shorter bouts overall with tighter dispersion, consistent with their faster ultradian REMS--NREMS cycling. Nevertheless, the overall increase in both the scatter trends and binned medians indicates that $P(t,\Delta)$ captures a conserved aspect of REMS cycle organization: the state of the system at REMS entry, {as summarized by REMS propensity, predicts the duration of the imminent REMS episode.}}

\allblack
\subsection{Temporal organization of REMS expression across the human sleep episode}
{Previous work has shown that REMS in humans is under strong circadian regulation and increases across the nighttime sleep episode \cite{czeisler1980, dijkczeisler1995, dijk2010age}.
To understand how these observations interact with the occurrence of single and sequential REMS cycles, we provide a detailed characterization of the temporal organization of REMS expression in our human data. First, we identified normalized sleep-episode deciles  for each sleep record (0–100\% of total time spent in sleep by the subject in increments of 10\%). 
This normalization accounts for different total sleep durations across sleep records }and allows us to determine whether REMS features, such as number of REMS bouts, REMS bout durations, sequential versus single REMS cycles, and overall time in REMS, are uniform across the sleep episode or exhibit structured temporal gradients. Such gradients would be expected if the underlying regulatory processes governing REMS expression, including circadian REMS modulation and the homeostatic dissipation of NREMS pressure, evolve systematically over the course of the night. 
The following subsections demonstrate that REMS expression in humans is not temporally uniform but shows marked increases at the very beginning and at the very end of the sleep episode. 

\begin{figure}[ht!]
    \centering
    \includegraphics[width=1\linewidth]{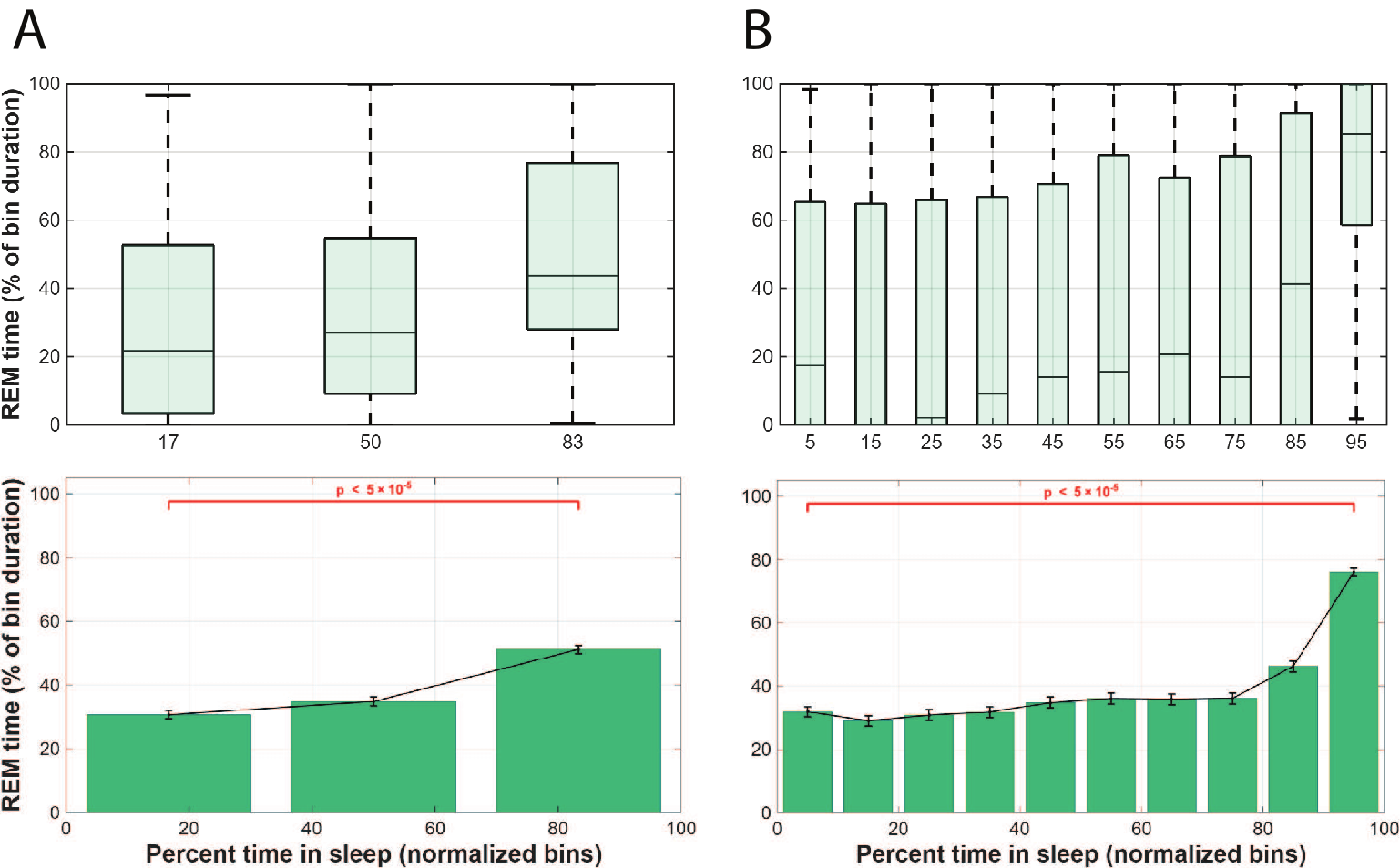}
    \caption{\textbf{Percent time in REMS increases toward the end of the normalized sleep episode in humans.}
Coarse three-bin summary (A) and fine scale ten-bin summary (B) of the fraction of time spent in REMS across the normalized sleep episode after long-wake exclusion and subject-level cycle-count IQR filtering.  Top: box-and-whisker plots across subjects for the subject-level REMS fraction in each normalized time bin.  Bottom: bars show subject means and error bars show standard error. (A): A subject-level one-sided sign-flip test comparing the last bin with the first bin showed a significant increase in REMS percentage toward the end of sleep (\(p < 5\times10^{-5}\)). 
(B): The final decile exhibits a marked increase in REMS percentage relative to the beginning of sleep; a subject-level one-sided sign-flip test comparing the last decile with the first decile was highly significant (\(p < 5\times10^{-5}\)). In all panels, whiskers indicate the most extreme observations within \(1.5\times\mathrm{IQR}\), and the REMS fraction is expressed as the percent of bin duration occupied by REMS. }
    \label{fig:percent_time_bins}
\end{figure}

\paragraph{Percent time spent in REMS increases across the night.}
Similar to previous reports \cite{dijk2010age}, we find that the percent time spent in REMS increases across the sleep episode. Figure~\ref{fig:percent_time_bins} shows the percent time spent in REMS when the sleep episode is divided into three (A) or ten (B) equal sized bins (bin duration varies with length of a subject's sleep episode).  While the coarse three-bin representation shows a monotonic increase similar to previous results \cite{dijk2010age}, the finer ten-decile version reveals the predominance of REMS in the very late stages of the sleep episode. The box-and-whisker plots (Figure~\ref{fig:percent_time_bins}, top row) display the high variability in the data while the bar plots (Figure~\ref{fig:percent_time_bins}, bottom row) show trends in the means. While the first tenth of the sleep episode contains a similar fraction of REMS as occurs during the middle of the sleep episode, REMS fraction is not uniform in the intervening deciles. At the very end of the sleep episode, REMS percentage dominates with REMS fraction reaching up to 70\% in the last decile.  

{This increase in REMS percent toward the end of sleep was supported statistically by subject-level one-sided sign-flip permutation tests (see Supplementary~\ref{sec: supp_fig_9} for more details): in the three-bin analysis, the last bin showed a significant increase in REMS percentage relative to the first bin, and in the decile analysis, the final decile was likewise significantly greater than the first decile (both $p < 5 \times 10^{-5}$). 
The results indicate that the late-night predominance of REMS is a robust population-level feature, and they support the interpretation that REMS expression becomes increasingly favored as the sleep episode progresses. 
At the same time, the elevated REMS fraction in the first decile suggests that REMS temporal organization is not purely monotonic, but instead reflects a more structured within--night pattern with enhanced REMS expression at {the beginning and end} of the sleep episode. }

\begin{figure}[H]
    \centering
    \includegraphics[width=1\linewidth]{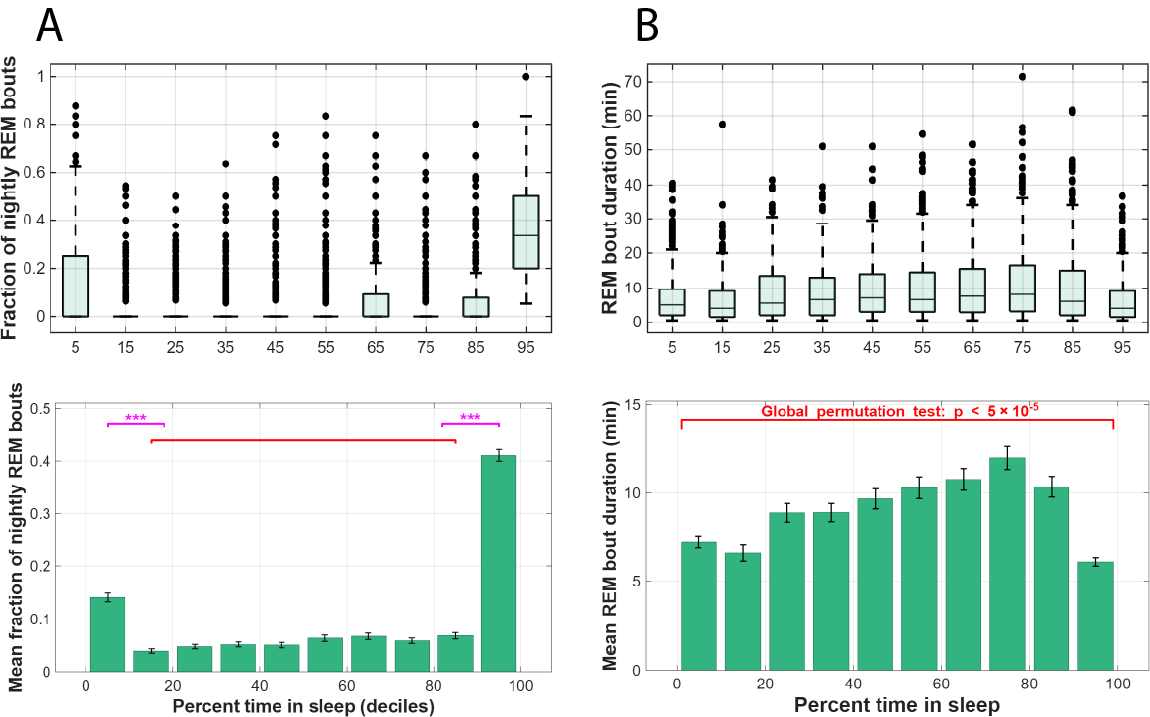}
\caption{\textbf{Normalized-time distribution of REMS onset frequency and REMS bout duration across the human sleep episode.}
Human REMS cycles were analyzed after long-wake exclusion (\(\geq 2\) min) and subject-level cycle-count IQR filtering by mapping each subject’s sleep episode onto a normalized time axis and dividing it into deciles. 
(A) REMS onset distribution across normalized sleep time. Top: box-and-whisker plots show the subject-level distribution of the fraction of nightly REMS bouts whose onsets occur in each decile. Bottom: bars show subject means and error bars show standard error across subjects. REMS onsets were strongly enriched near {the beginning and end} of the sleep episode, especially in the final decile. The red bracket denotes the middle deciles (10--90\% of normalized sleep time), and the magenta asterisks indicate significant edge-versus-middle comparisons based on subject-level paired sign-flip permutation tests: the first decile differed significantly from the middle deciles (\(p < 5\times10^{-5}\)), and the last decile also differed significantly from the middle deciles (\(p < 5\times10^{-5}\)). 
(B) REMS bout duration as a function of normalized onset decile. Top: box-and-whisker plots show the pooled distribution of REMS bout durations for bouts whose onsets fell in each decile. Bottom: bars show pooled mean REMS bout duration in each decile and error bars show standard error across bouts. REMS bout duration varied significantly across onset deciles, with the longest bouts occurring in the later-middle portion of the sleep episode and shorter bouts at {the beginning and end}, particularly in the final decile. The red bracket indicates the global across-decile comparison, and a within-subject label permutation test confirmed a significant dependence of REMS bout duration on onset decile (\(p < 5\times10^{-5}\)). In all box-and-whisker plots, whiskers extend to the most extreme observations within \(1.5\times \mathrm{IQR}\), black markers indicate outliers.}
    \label{fig:rem_distributions}
\end{figure}

\paragraph{Number and duration of REMS bouts across the sleep episode.}
To further quantify how REMS varies across the sleep episode, Figure~\ref{fig:rem_distributions} shows the mean fraction of nightly REMS bouts whose onsets occur in each normalized-time decile (A) and the mean REMS bout duration (min) for bouts initiated in each decile (B). Across subjects, REMS onset events were strongly concentrated near {the beginning and end} of the sleep episode: approximately 15\% of nightly REMS bouts were initiated in the first decile, whereas about 41\% were initiated in the last decile. Subject-level paired sign-flip permutation tests confirmed that the first decile, the last decile, and the average of the two edge deciles each differed significantly from the middle deciles (10--90\% of normalized sleep time; all \(p<5\times10^{-5}\); see Supplementary~\ref{sec: supp_fig_9} for details of the hypothesis tests). 

REMS bout duration also varied significantly across onset deciles (within-subject label permutation test, \(p<5\times10^{-5}\); see Supplementary \ref{sec: supp_fig_9}), but its pattern was not monotone. REMS bout durations were relatively short in the earliest and latest deciles, increased across the middle portion of the sleep episode, and reached their largest values in the later-middle deciles before declining again near the end of sleep. Thus, the elevated REMS percentages near the end of the sleep episode shown in Figure~\ref{fig:percent_time_bins} appear to be driven by increasing REMS duration until the very end of the sleep episode when shorter, more frequent REMS bouts occur. 

\paragraph{Single and Sequential REMS cycles exhibit distinct temporal profiles.} 
To further dissect the temporal structure of REMS architecture, we examined the organization of REMS into single and sequential cycles as a function of REMS onset time within the normalized sleep episode. 
As shown above (Figure~\ref{fig:rem_distributions}) the distribution of REMS onsets across the night is highly non-uniform with a substantial fraction of nightly REMS bouts occurring early and late in sleep. 
Here, we ask how this overall temporal distribution decomposes into single versus sequential REMS organization, and whether these two REMS modes differ in their temporal evolution. 

\begin{figure}[ht!]
    \centering
    \includegraphics[width=1\linewidth]{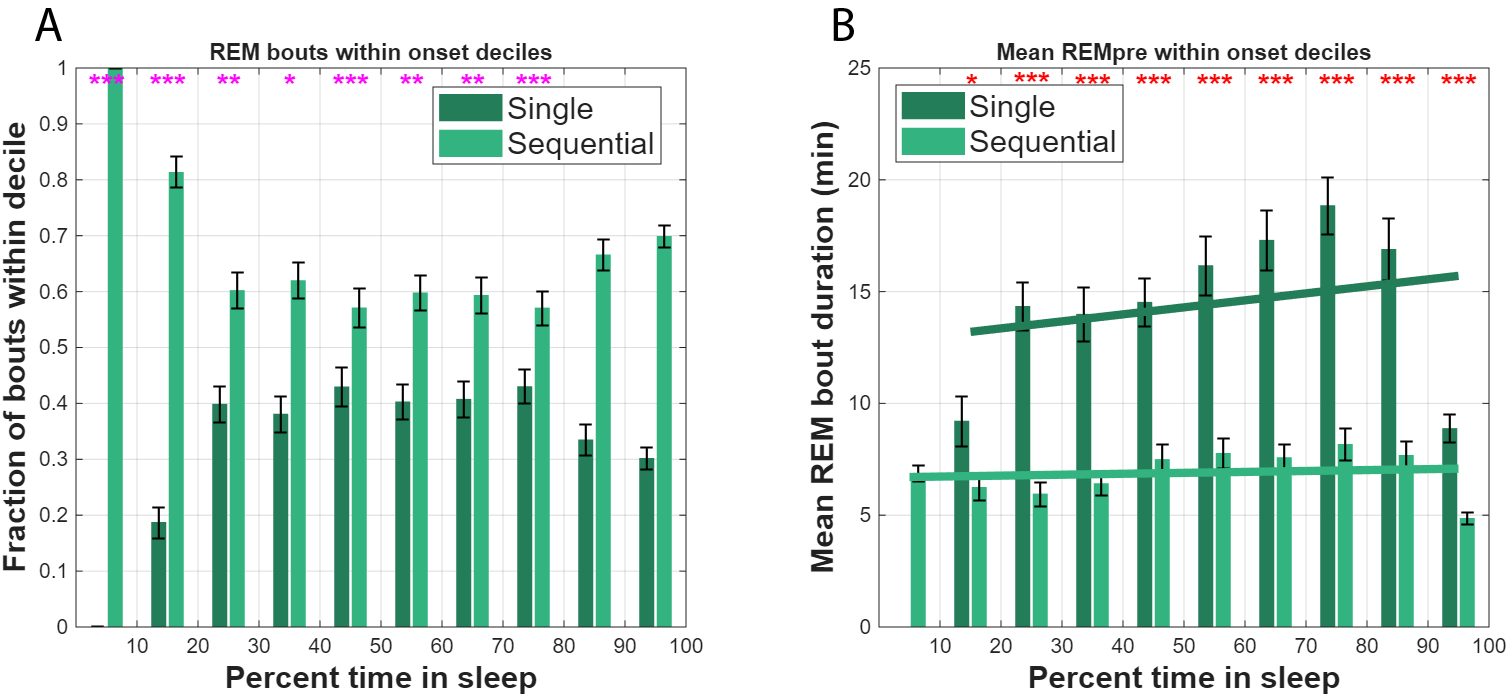}
\caption{
{\bf Single vs.\ sequential REMS cycle occurrence across the sleep episode.}
REMS cycles were classified as single and sequential. Each cycle was assigned to a decile based on REMS onset time within a normalized sleep episode.
(A) Pooled fraction of REMS bouts with onset in each decile that initiate single (dark green) or sequential (light green) REMS cycles. Error bars denote binomial standard errors. Magenta asterisks indicate deciles in which the fraction of single REMS cycles differs significantly from the overall pooled baseline (two-sided exact binomial test, false discovery rate corrected), with one, two, and three asterisks denoting \(p_{\mathrm{FDR}}<0.05\), \(p_{\mathrm{FDR}}<0.01\), and \(p_{\mathrm{FDR}}<0.001\), respectively.
(B) Pooled mean REMS bout duration (\rempre, min) within each onset decile in single (dark green) or sequential (light green) REMS cycles. Error bars denote standard errors of the mean. Red asterisks indicate deciles with significant differences in REMS bout duration between single and sequential cycles (Welch two-sample \(t\)-test, false discovery rate corrected), using the same significance thresholds. Global within-subject permutation tests further showed that REMS bout duration depended significantly on onset decile for both single cycles (\(p<5\times10^{-5}\)) and sequential cycles (\(p=0.0115\)).
{The fitted trend lines further show these distinct temporal patterns: mean duration of REMS bouts in single cycles shows a clear positive trend across onset deciles, increasing from early to later portions of the sleep episode, whereas mean duration of REMS bouts in sequential cycles remains comparatively flat, showing only a weak change across deciles.}}
    \label{fig:seq_single}
\end{figure}

Figure~\ref{fig:seq_single}A shows the composition of REMS cycles within each REMS-onset decile, expressed as the pooled fraction of cycles classified as single or sequential. 
Consistent with the distribution of REMS onsets across the sleep episode (Figure~\ref{fig:rem_distributions}A), early REMS expression is qualitatively distinct: among the REMS cycles initiated in the first decile, the overwhelming majority are classified as sequential rather than single. 
This bias indicates that early REMS onsets predominantly occur in clustered REMS--NREMS--REMS sequences separated with short inter-REMS intervals. 
As the sleep episode progresses into mid-sleep, the balance shifts significantly, with single REMS cycles contributing more to REMS expression, reflecting longer inter-REMS accumulation between REMS bouts. At the very end of the sleep episode, the occurrence of sequential REMS cycles increases again, indicating a partial re-emergence of interrupted REMS organization in late sleep.

Figure~\ref{fig:seq_single}B shows the mean REMS bout duration (\rempre) for single and sequential cycles within each REMS-onset decile.
Across the sleep episode, REMS bouts in single cycles tend to exhibit longer mean durations than REMS bouts in sequential cycles, indicating that these two REMS cycling regimes differ not only in their temporal organization but also in REMS bout length.
This separation is statistically supported by two-sample Welch t-tests comparing \rempre \ between single and sequential cycles within each decile: from the second decile onward (approximately 20–90\% of the sleep episode), REMS bouts in single cycles are significantly longer than in sequential cycles after false discovery rate correction ($p_{\mathrm{FDR}}<0.05$).
No significant difference was detected in the earliest decile, where single REMS cycles are rare.

Beyond this categorical distinction, REMS bouts in single cycles exhibit a pronounced modulation across the sleep episode, increasing in duration from early to mid-sleep and reaching their longest durations in the latter half of the sleep episode.
In contrast, REMS bouts in sequential cycles remain comparatively short throughout the sleep episode, showing only modest lengthening toward the end of sleep.


{This analysis of normalized sleep-time deciles shows that REMS expression in humans is strongly time-dependent across the sleep episode. 
The fraction of time spent in REMS increases toward the end of sleep, with the most pronounced rise occurring in the final decile, where REMS occupied the largest share of bin time (Figure~\ref{fig:percent_time_bins}), indicating that REMS becomes progressively more dominant as the night advances. 
REMS bout onsets were also highly non-uniform across the episode, being enriched near both {the beginning and end} of sleep and especially in the final decile (Figure~\ref{fig:rem_distributions}A), suggesting that late REMS predominance is driven not only by increased REMS occupancy but also by a greater likelihood of entering REMS near the end of the sleep episode. 
In contrast, REMS bout duration did not increase monotonically across the night: bouts were shorter at the beginning and end of the sleep episode and longest in the later-middle deciles (Figure~\ref{fig:rem_distributions}B). 
This indicates that the large REMS fraction late in sleep is not simply due to longer REMS bouts, but instead reflects more frequent REMS initiation, with the final part of sleep characterized by shorter but more frequent REMS episodes.
Decomposing REMS expression into sequential and single cycles revealed distinct temporal structure: early REMS was dominated by sequential cycles, mid-sleep REMS was dominated by single cycles, and late sleep showed a renewed increase in sequential REMS cycle occurrence (Figure~\ref{fig:seq_single}A). 
This pattern suggests that REMS organization changes systematically across the night, with early and late sleep showing more interrupted or clustered REMS expression, and mid-sleep showing more consolidated REMS expression separated by longer time in NREMS.
Across most onset deciles, REMS bouts in single cycles were significantly longer than the bouts in sequential cycles (Figure~\ref{fig:seq_single}B), indicating that temporal changes in REMS timing across the sleep episode are accompanied by systematic shifts in REMS micro-architecture.
Together, these findings may reflect an evolving interaction between homeostatic and circadian influences: early in the night, strong homeostatic NREMS pressure may limit REMS consolidation, whereas later in the night, dissipation of NREMS pressure together with rising circadian REMS drive may promote frequent REMS initiation and a distinct late-sleep REMS organization (see Discussion). 
}

\allblack
\section{Discussion}

In this study, we analyzed human, rat, and mouse ultradian NREMS-REMS cycle data using the same methodology and found that the ultradian cycle structures showed many similarities among the 3 species despite differences in time scales and diurnal vs nocturnal behavior. In all 3 species, REMS bout durations were positively correlated with the subsequent inter-REMS interval duration such that longer REMS bouts were followed by longer inter-REMS intervals. This result supports the hypothesis of an hourglass-type homeostatic process driving REMS sleep in which a longer REMS episode discharges a greater portion of accumulated REMS pressure, thereby delaying the onset of the next REMS episode \cite{beningtonheller1994, beningtonheller1994review}. Furthermore, this result is not  inconsistent with the hypothesis of a post-REMS refractory period which occurs after REMS and inhibits the initiation of a subsequent REMS bout {for a period of time }proportional to the REMS episode \cite{le2021asymmetrical}.   

Distributions of the duration of NREMS in inter-REMS intervals ($|N|$) were right-skewed for all 3 species. The human data showed a bimodal profile as previously reported \cite{zamboni1999control,esposito2003,esposito2004} and was fit well with a three-part mixture model that accounted for the high density at the lowest scoring epoch of 30 s, an initially slowly decaying tail and an approximately normal distribution at longer durations.  {In the rodent data, bimodality was more clearly revealed in the $\log(|N|)$ distribution and when REMS cycles were partitioned by REMS duration consistent with previous work  \cite{park2021probabilistic, ginsberg2024predictive}.} For both mouse and rat data, these distibutions were fit well with two-component Gaussian mixture models. {The short- and long-components of these mixture models correspond to  sequential and single REMS cycles, respectively, and the intersection between these components provides a quantitative method to identify the threshold separating sequential and single REMS cycles.}  
In both mouse and rat data, there was less separation between sequential and single REMS cycles for the shortest \rempre \ bins, reflecting that the majority of those inter-REMS intervals were short. For the longer \rempre \ bins, the long model component generally shifted to longer $|N|$ values with increasing \rempre\ as predicted by the positive correlation between durations of REMS bouts and the subsequent inter-REMS interval.

Based on these mixture model fits for $|N|$ (human) or $\log(|N|)$ (rodent) distributions, we computed REMS propensity, $P(t, \Delta)$, as a function of time $t$ spent in NREMS during the inter--REMS interval. {Although different mixture models were used for different species, we} found that REMS propensity functions had similar profiles across species, with minimum propensity values occurring for shorter $|N|$  and increasing to a local peak as $|N|$ increased. Then, the propensity decreased for the remaining longer values of $|N|$. In the rodent data, the $|N|$ value at propensity peak increased for longer durations of \rempre\ (albeit with some variability in this trend for the rat propensity), reflecting that longer REMS bouts are followed by longer inter-REMS intervals. This trend was not apparent in the human propensity but this may be due to only considering a small number of \rempre\ bins.

Importantly, the REMS propensity measure at REMS onset predicted REMS bout duration in all 3 species we considered.  {Specifically, in humans and rats, REMS bout duration was positively correlated with REMS propensity at REMS bout onset, consistent with previous work in mice \cite{ginsberg2024predictive}.} This finding highlights the predictive value of the REMS propensity measure, $P(t, \Delta)$, to represent REMS pressure based on the amount of accumulated NREMS since the previous REMS bout. Furthermore, it provides a mechanism to begin understanding both the timing and duration of REMS episodes. 

While short REMS bouts and short inter-REMS intervals are a widely acknowledged feature of rodent polyphasic sleep, standard sleep scoring practices for human data have tended to ignore short interruptions of sleep states in favor of scoring consolidated bouts of NREMS and REMS as they alternate across the night \cite{charles1980glossary, merica1991study}. However, raw EEG recordings of human sleep indicate a more complex story, and this scoring practice may mask similarities in the temporal architecture of sleep in different species. {Although NREMS and REMS typically alternate on a 90 minute ultradian timescale in human sleep, the``REM sleep" part of the cycle may include either longer, consolidated REMS bouts or shorter, fragmented bouts of REMS interspersed with NREMS. When scored without criteria to promote consolidation, the alternation between periods of majority REMS and majority NREMS on a 90 minute timescale is preserved, but many shorter REMS-NREMS cycles are observed.} As we seek to further our understanding of the mechanisms driving REMS-NREMS ultradian cycling, these fine details of sleep architecture may be important considerations. 

Previous work to distinguish REMS cycles based on the duration of associated inter-REMS intervals was quantitatively formalized in rats as single (one longer bout) or sequential (multiple shorter bouts close together) REMS cycles \cite{zamboni1999control}. These dynamics have also been observed in human sleep data \cite{merica1991study, esposito2003}, however, to our knowledge, they have not been quantitatively formalized beyond being used to designate an arbitrary threshold for inter-REMS intervals \cite{merica1991study}. {We formalized this analysis for a heterogeneous collection of human data representing participants from a range of demographic groups, collected over different time periods, with different inclusion/exclusion criteria, and with different experimental and scoring protocols. Despite the heterogeneity, a clear bimodal distribution in inter-REMS interval duration was observed in the pooled data. The inter-REMS NREMS duration ($|N|$) data were fit with a mixture model that, although it had different components compared to the mixture model for the rodents, provides a data-driven, quantitative method for differentiating sequential and single REMS cycles and computing REMS propensity.}

Furthermore, our analysis of the temporal organization of REMS bouts across the human sleep episode helps to better understand how single and sequential REMS cycles contribute to the well-documented increase in REMS across the sleep episode in humans. By considering REMS bouts within deciles of the normalized sleep episode, we found that a higher fraction of REMS bouts occurs within the first and last deciles of the sleep episode. Interestingly, the majority of REMS cycles in the first decile were sequential while there was a higher proportion of single REMS cycles during mid-sleep. During the last deciles, sequential cycles again occurred more often than single cycles. {This suggests that at the beginning and end of a sleep episode, REMS bouts are more likely to be fragmented. This fragmentation may reflect effects of NREMS homeostasis and circadian modulation of REMS.} For example, early in the sleep episode, the predominant occurrence of sequential cycles with short REM bouts  may reflect unstable REMS expression under strong homeostatic NREMS pressure. During mid-sleep, REMS bouts become longer and more consolidated leading to dominance of single REMS organization with a lower fraction of REMS onsets, potentially caused by decreasing NREMS pressure and increasing circadian REMS drive. Late in the sleep episode, when the percent time spent in REMS is the highest, the reappearance of sequential REMS organization and the eventual decrease in REMS bout durations may reflect { high circadian REMS drive expressed in the absence of high NREMS pressure}.  

Our findings demonstrate that changes in REMS timing across sleep episodes in humans are accompanied by systematic shifts in REMS micro-architecture revealing distinct temporal regimes of REMS organization likely governed by evolving homeostatic and circadian influences. Future work utilizing controlled experiments with standardized scoring practices that allow for fragmentation of states are needed to understand this dynamic process. Furthermore, such studies may enable the investigation of age, sex, or disease effects on single and sequential REMS cycles as well as ultradian cycling.

As experimental studies continue to investigate the neural regions and processes that govern ultradian NREMS-REMS alternation in rodent models, it is important to account for how those processes influence single and sequential REMS cycles. {Moreover, as we work to infer whether the same neural processes govern human ultradian cycling, it will be important to understand the temporal architecture of single and sequential REMS cycles in humans. Our analysis identifies clear similarities in human and rodent ultradian cycling and REMS regulation while also formalizing some potential differences. Future work in these and other species is needed to identify the significance of single and sequential REMS cycles and the role of circadian modulation and homeostatic pressure for both REMS and NREMS in producing these cycles.}


\section{Methods}

\subsection{Sleep recordings}
\subsubsection{Human}
Human sleep recordings (573 total recordings) from subjects without serious health or sleep conditions were collected from 3 public online databases containing polysomnographic (PSG) recordings: (1) the Sleep-EDF Database Expanded on Physionet (\url{https://www.physionet.org/content/sleep-edfx/1.0.0/}), (2) Mignot Nature Communications dataset supported by the National Sleep Research Resource (\url{https://sleepdata.org/datasets/mnc)}, and (3) Bitbrain Open Access Sleep (BOAS) dataset (\url{https://www.bitbrain.com/science/eeg-datasets}).

From the Sleep-EDF Database Expanded, we used data from 152 subjects in the Sleep Cassette Study which includes PSG sleep records from healthy Caucasians ($54\%$ female) aged 25-101 who were not taking any sleep-related medication \cite{sleep_edf_kemp,sleep_edf_goldberger}. {{Each participant (except 3) underwent two subsequent nights of recordings in their own homes. This resulted in 153 recordings from 78 unique participants.}} We used the hypnogram files that contained annotations of sleep patterns corresponding to the PSG in 30-s epochs. {{The Sleep-EDF dataset was collected from 1987-1991 and scored based on the Rechtschaffen and Kales (R \& K) scoring criteria \cite{rechtschaffen1968manual}.}} Scored states were wake, REM sleep, NREM states 1, 2, 3 and 4, and movement. We renamed all NREM sleep states as a single NREM state. 

From the Mignot Nature Communications (MNC) dataset, we used data from 294 control (non-narcoleptic) subjects from three of the study cohorts (SSC, CNC and DHC) \cite{mnc_zhang, mnc_stephansen}.  {{From the MNC dataset, we selected three study cohorts: the Stanford Sleep Cohort (SSC) \cite{andlauer2013nocturnal, moore2014design}, the Chinese Narcolepsy Cohort (CNC) \cite{andlauer2013nocturnal}, and the Danish Hypersomnia Cohort (DHC) \cite{christensen2017novel}. The SSC cohort recordings were scored using the R \& K critera \cite{moore2014design, rechtschaffen1968manual}. The PSG recording of the DHC participants were scored by experienced technicians using the American Academy of Sleep Medicine (AASM) criteria \cite{berry2015aasm}. }} The data in these cohorts was annotated in 30-s epochs as wake, REM sleep, NREM sleep states 1, 2 and 3, and unscored. We renamed all NREM sleep states as a single NREM state.

\begin{figure}[H]
    \centering
    \includegraphics[width=0.6\linewidth]{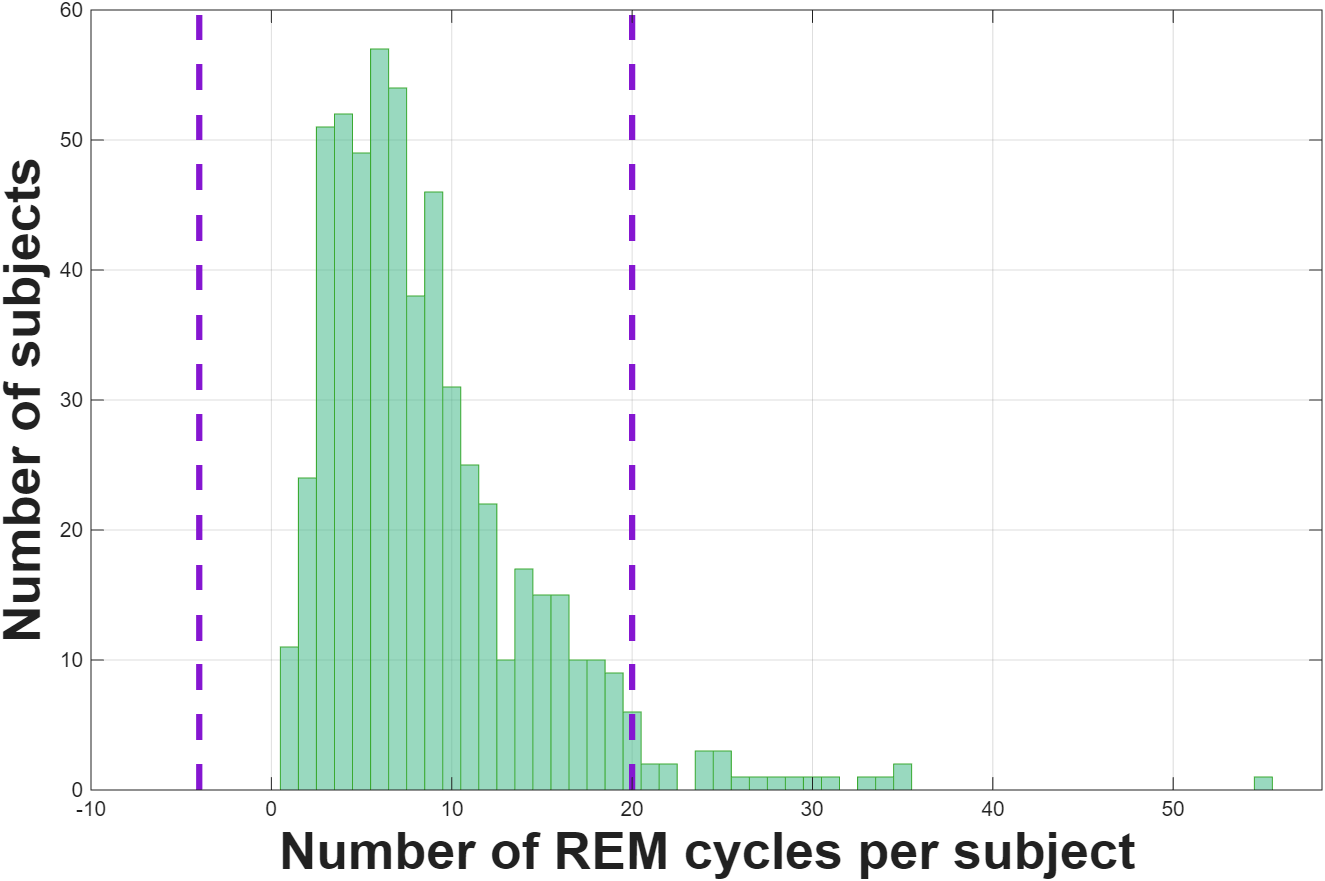}
    \caption{Subject-level distribution of retained REMS cycle counts in the human dataset after application of the long-wake filtering criterion. Dashed vertical lines denote the quartile-based outlier bounds, $Q_1-1.5\,\mathrm{IQR}$ and $Q_3+1.5\,\mathrm{IQR}$, used to remove subjects with atypical numbers of analyzable REMS cycles. This subject-level filtering step was applied before human mixture-model fitting.}
    \label{fig:placeholder}
\end{figure}

The BOAS dataset contains PSG recordings from  participants who ranged in age (18-82 years old), sex ($59\%$ female), and body mass index (18.3 - 33.3) \cite{bitbrain_dataset}. There were 108 unique individuals who participated in the study. Sleep staging was scored by three experts who independently scored the data following criteria developed by the American Academy of Sleep Medicine (AASM) \cite{berry2015aasm}. Then, a consensus label was determined \cite{bitbrain_dataset}. Exclusion criteria only included severe conditions that may have affected the feasibility or safety of the protocol. We used data from 127 recordings that were manually annotated in 30-s epochs as wake, REM sleep, NREM sleep states 1, 2 and 3. We renamed all NREM sleep states as a single NREM state. 

{
Because the human dataset was assembled from multiple public cohorts and included recordings of varying duration and structure, we applied an additional subject-level outlier filter to limit disproportionate influence from subjects contributing an unusually small or unusually large number of analyzable REMS cycles (Figure \ref{fig:placeholder}). After applying the long-wake (LW) exclusion criterion (see below) and retaining only valid REMS cycles, we counted the number of remaining cycles contributed by each subject. Letting $Q_1$ and $Q_3$ denote the first and third quartiles of the subject-wise cycle counts, and $\mathrm{IQR}=Q_3-Q_1$, we classified as outliers any subjects with cycle counts below $Q_1-1.5\,\mathrm{IQR}$ or above $Q_3+1.5\,\mathrm{IQR}$. All REMS cycles from such subjects were excluded from our analysis. This procedure was designed to reduce undue influence from atypical recordings while preserving the cycle-level distribution within the typical range of subject contributions.}
After this filtering step, 515 of the 573 human recordings remained for analysis.

\subsubsection{Rat}
Rat sleep data were collected from control condition experiments from \cite{silverstein_ratdata}. Briefly, adult Sprague Dawley rats (n=5 rats (4 female, 1 male), 300-350 g,  aged 8-12 weeks,  Charles River Inc., Wilmington, MA) maintained on a 12:12 light:dark cycle (lights on at 8:00am) and with ad libitum food and water, were used for all recordings. The experimental protocol was approved by the Institutional Animal Care and Use Committee at the University of Michigan, Ann Arbor, and was conducted in compliance with the Guide for the Care and Use of Laboratory Animals (Ed 8, National Academies Press) and ARRIVE Guidelines. Under isoflurane anesthesia, stainless steel screw electrodes were implanted across the cortex for electroencephalogram (EEG) recordings and wires were positioned in the dorsal nuchal muscles to record electromyogram (EMG). Monopolar EEG (0.1-300 Hz, sampling rate 1 kHz) was recorded in a bipolar montage (0.1-125 Hz, sampling rate 500 Hz) for polysomnography (frontal-frontal, frontal-parietal, and parietal-parietal). The EMG was bandpass filtered between 0.1 Hz and 125 Hz and sampled at 500 Hz. Bipolar EEG and EMG data over a 48-h period were manually scored (SleepSign; Kissei Comtec Inc., Matsumoto, Japan) in 4-s epochs into 1) wake state: low-amplitude fast EEG along with high muscle tone, 2) NREMS: high-amplitude slow EEG along with low muscle tone, and 3) REMS: low-amplitude fast EEG along with muscle atonia. 

\subsubsection{Mouse}
Mouse sleep data were collected as described in \cite{park2021probabilistic}. Briefly, male or female C57BL/6J mice (Jackson Laboratory stock no. 000664) housed on a 12-h dark/12-h light cycle (lights on between 7 am and 7 pm), aged 6–12 weeks with ad libitum access to food and water were used for all recordings. All experimental procedures were approved by the Institutional Animal
Care and Use Committee (IACUC) at the University of Pennsylvania and conducted in accordance with the National Institutes of Health Office of Laboratory Animal Welfare Policy. Under isoflurane anesthesia, stainless steel screw electrodes were implanted over the parietal and prefrontal cortex and the cerebellum for EEG recordings, and wires were inserted into the neck muscle for EMG recording. EEG and EMG signals, recorded using an RHD2000 amplifier (intan, sampling rate
1 kHz), were processed by custom software in 2.5-s epochs into 1) wake state: low delta (0.5-4 Hz) power and/or high gamma (100-150 Hz) and high EMG power; 2) NREMS: high delta power,low theta (5-12 Hz) /delta power ratio and low EMG power; and 3) REMS: high theta/
delta power ratio, low EMG power, and low delta power. All recordings were then manually rescored to verify classification. Data are available online at \url{https://zenodo.org/records/5817119\#.YdvvNC\_kGTc} and \url{https://zenodo.org/records/5820559\#.Ydvvcy\_kGTc}.

\subsection{Computation of REMS propensity} 

REMS propensity is quantified as a predictive conditional probability of entering REMS based on the local sleep history preceding each REMS bout. Our analysis builds on the probabilistic framework introduced in previous work \cite{park2021probabilistic, ginsberg2024predictive} and we summarize the key steps here. 

\paragraph{REMS cycle definition, \rempre, \rempost, and \rempre \ binning.}
Sleep was partitioned into REMS cycles, defined as the interval from the onset of
one REMS bout to the onset of the next.
Each cycle is characterized by the duration of its initial REMS bout
(\rempre), followed by an inter-REMS interval of duration \irem \ with cumulative time spent in NREMS sleep $|N|$.
The REMS bout terminating the cycle is referred to as REMpost and represents
the subsequent expression of REMS following the inter-REMS interval.
Thus, each REMS cycle captures the relationship between an initial REMS episode
(\rempre), the intervening inter-REMS interval, and the timing and duration of the next REMS bout (\rempost).


\paragraph{Long wake filtering and statistical validation.}
To ensure that REMS propensity reflects intrinsic inter-REMS dynamics during sleep behavior, we excluded REMS cycles containing prolonged wake episodes within the inter-REMS interval. 
The choice of the threshold defining a long wake (LW) episode was statistically validated using two-sample Kolmogorov-Smirnov (KS) tests and Bayesian Information Critierion (BIC) tests {(see Supplementary~\ref{sec: supp_filter})}. 
Specifically, for candidate wake duration thresholds, REMS cycles were separated into those with and without wake bouts exceeding the threshold, and the resulting distributions of cumulative inter-REMS $|N|$ quantities were compared. 
KS and BIC tests revealed more stable, well-structured inter-REMS $|N|$ distributions when cycles containing sufficiently long wake episodes were removed. 
Based on this validation, an LW threshold of $\geq 2$ min was used for mouse, rat and human sleep. 

\paragraph{Mixture model fitting and goodness-of-fit.}
REMS cycles were partitioned into predefined \rempre \ bins of 30 s intervals for rodent data and into 3 bins with equal numbers of REMS cycles for human data.
For REMS cycles in each \rempre\ bin, the distributions of cumulative NREMS duration in the inter-REMS interval ($|N|$) were fit with mixture models. For rodent data, distributions of $\log(|N|)$ were fit with Gaussian mixture models (GMMs) as in previous work \cite{park2021probabilistic,ginsberg2024predictive}.  For human data, $|N|$ distributions were fit with a three-part mixture model consisting of an atom ( point mass) at $x_{\min}$, a short-duration continuous component given by the normalized $E1$ form $\propto e^{-rt}/t$ on $[x_{\min}, \infty)$, and a long-duration truncated normal component (see Supplementary \ref{sec: supp_human_fit}).
Model adequacy was assessed using corrected Kolmogorov-Smirnov tests to ensure consistency between the fitted models and empirical distributions (for rodent data, see Supplementary Tables \ref{tab:ks_distances_ksbest_bins_by_species} and \ref{tab:ks_pvalues_bins_by_species}; for human data, see Supplementary \ref{sec: supp_human_fit}).

\paragraph{Definition of single and sequential REMS bouts.}
{
In the mixture model fits of the distributions of inter-REMS NREMS accumulation $|N|$ in each \rempre\ bin, the two mixture components correspond to a short-$|N|$ mode and a long-$|N|$ mode. 
We defined the boundary between regimes as the intersection point of the two weighted component density curves, i.e., the value of $|N|$ at which the probabilities of belonging to the short and long modes are equal. 
Cycles with $|N|$ below the intersection (more likely to belong to the short-$|N|$ mode) were classified as sequential REMS cycles, reflecting REMS cycles with short inter-REMS NREMS duration.  
Cycles with $|N|$ above the intersection (more likely to belong to the long-$|N|$ mode) were classified as single REMS cycles, corresponding to REMS episodes followed by more substantial NREMS accumulation before the next REMS episode.
}

\paragraph{REMS propensity function.}
Following \cite{ginsberg2024predictive}, REMS propensity was defined as the conditional probability that a transition into REMS occurs before an additional fixed increment of NREMS accumulates. 
This quantity can be interpreted as a discrete time hazard-like function derived from the cumulative distribution function of inter-REMS NREMS duration $|N|$.
Propensity functions were computed within each \rempre\ bin and evaluated at REMS onset, yielding a predictive measure of local REMS sleep pressure.

\subsection{Analysis of REMS expression across the normalized human sleep episode}

To characterize how REMS expression varies across the human sleep episode, we analyzed the temporal distribution of REMS in the human data after applying the LW exclusion criterion (see above) and a subject-level outlier filter based on the number of valid REMS cycles exhibited per subject after LW filtering. Sleep records with cycle counts outside the interquartile-range criterion ($< Q_1-1.5\,\mathrm{IQR}$ or $> Q_3+1.5\,\mathrm{IQR}$) were excluded to reduce disproportionate influence from outliers. Only retained REMS cycles were used in our analyses.

For each subject, the duration of the complete sleep episode was mapped onto a normalized time axis spanning 0--100\% which was then partitioned into ten equal deciles. The normalized time of onset for each of the subject's retained REMS bouts was then computed by mapping REMS episode onset times from the start of the sleep episode onto the normalized time axis. Normalized REMS bout onset times were then assigned to their appropriate deciles. 

After normalizing REMS bout onset times for all subjects, we then quantified three complementary features of REMS expression across normalized time in sleep.
First, to measure REMS onset distribution, we computed for each subject the fraction of REMS bouts whose onsets occurred in each decile. 
Second, to quantify the distribution of REMS time across the sleep episode, we calculated for each subject the fraction of total REMS time falling within each decile. 
For this analysis, REMS time was assigned to deciles according to its temporal overlap with each decile bin.
Third, to determine how REMS bout duration depended on its onset time in the sleep episode, for each decile we computed the mean durations of REMS bouts whose onset occurred in that decile, regardless of whether the REMS bout duration temporally overlapped with neighboring deciles. 

For the REMS onset-distribution and REMS-time analyses, subject-level fractions were first computed within each subject and then averaged across subjects. For the bout-duration analysis, descriptive plots show pooled bout-level means by decile, whereas statistical inference was performed with within-subject permutation procedures to avoid pseudo-replication from treating multiple bouts from the same subject as independent observations.

\section*{Conflict of interest statement}
The authors declare that the research was conducted in the
absence of any commercial or financial relationships that could be construed as a potential conflict of interest.
\section*{Author Contributions}
Conceptualization (AGG, CGDB, VB), Methodology (AGG), Software (AGG, NA, MECC, YY), Formal Analysis (AGG, NA, MECC, YY), Investigation (FW, DP), Writing-Original Draft (NA, VB, SRS), Writing-Review and Editing (NA, AGG, MECC, YY, SRS, DP, FW, CGDB, VB), Supervision (CGDB, VB).

\section*{Funding}
This work was supported by the NIH National Center for Complementary and Integrative Health R01AT013188 (NA, SRS, CGDB, FW, VB) and National Institute of General Medical Sciences R01GM121919 (DP).


\section*{Supplementary Material}

Matlab codes used to generate results figures can be found online at \url{https://github.com/Naghmeh-Akhavan/REM-Propensity-Across-Multiple-Species.git}{}.

\bibliographystyle{ieeetr}
\bibliography{References.bib}

@article{czeisler1980,
  title={Human sleep: its duration and organization depend on its circadian phase},
  author={Czeisler, C. A. and Weitzman, E.D and Moore-Ede, M. C. and Zimmerman, J. C. and Knauer, R. S.},
  journal={Science},
  volume={210},
  number={4475},
  pages={1264--1267},
  year={1980}
}

@article{dijkczeisler1995,
  title={Contribution of the circadian pacemaker and the sleep homeostat to sleep propensity, sleep structure, electroencephalographic slow waves, and sleep spindle activity in humans},
  author={Dijk, D. J. and Czeisler, C. A.},
  journal={J Neurosci},
  volume={15},
  number={5},
  pages={3526--3538},
  year={1995}
}

@article{park2021probabilistic,
  title={A probabilistic model for the ultradian timing of {REM} sleep in mice},
  author={Park, Sung-Ho and Baik, Justin and Hong, Jiso and Antila, Hanna and Kurland, Benjamin and Chung, Shinjae and Weber, Franz},
  journal={PLoS computational biology},
  volume={17},
  number={8},
  pages={e1009316},
  year={2021},
  publisher={Public Library of Science San Francisco, CA USA}
}

@article{ginsberg2024predictive,
  title={A predictive propensity measure to enter {REM} sleep},
  author={Ginsberg, Alexander G and Cruz, Madelyn Esther C and Weber, Franz and Booth, Victoria and Diniz Behn, Cecilia G},
  journal={Frontiers in Neuroscience},
  volume={18},
  pages={1431407},
  year={2024},
  publisher={Frontiers Media SA}
}

@article{zamboni1999control,
  title={Control of {REM} sleep: an aspect of the regulation of physiological homeostasis.},
  author={Zamboni, G and Perez, E and Amici, R and Jones, CA and Parmeggiani, PL},
  journal={Archives italiennes de biologie},
  volume={137},
  number={4},
  pages={249--262},
  year={1999}
}

@article{vivaldi2005short,
  title={Short-term homeostasis of {REM} sleep throughout a 12: 12 light: dark schedule in the rat},
  author={Vivaldi, Ennio A and Ocampo-Garc{\'e}s, Adri{\'a}n and Villegas, Rodrigo},
  journal={Sleep},
  volume={28},
  number={8},
  pages={931--943},
  year={2005},
  publisher={Oxford University Press}
}

@article{vivaldi1994short,
  title={Short-term homeostasis of active sleep and the architecture of sleep in the rat},
  author={Vivaldi, ENNIO A and Ocampo, ADRIAN and Wyneken, URSULA and Roncagliolo, M and Zapata, AM},
  journal={Journal of neurophysiology},
  volume={72},
  number={4},
  pages={1745--1755},
  year={1994}
}

@article{le2021asymmetrical,
  title={An asymmetrical hypothesis for the {NREM}-{REM} sleep alternation—what is the {NREM}-{REM} cycle?},
  author={Le Bon, Olivier},
  journal={Frontiers in Neuroscience},
  volume={15},
  pages={627193},
  year={2021},
  publisher={Frontiers Media SA}
}

@article{benington1994remdep,
  title={{REM}-sleep propensity accumulates during 2-h {REM}-sleep deprivation in the rest period in rats},
  author={Benington, Joel H and Woudenberg, M Catherine and Heller, H Craig},
  journal={Neuroscience letters},
  volume={180},
  number={1},
  pages={76--80},
  year={1994},
  publisher={Elsevier}
}

@article{beningtonheller1994,
  title={{REM}-sleep timing is controlled homeostatically by accumulation of {REM}-sleep propensity during non-{REM} sleep},
  author={Benington, Joel H and Heller, H Craig},
  journal={Am J Physiol},
  volume={266},
  pages={R1992--R2000},
  year={1994},
}

@article{beningtonheller1994review,
  title={Does the function of {REM} sleep concern non-{REM} sleep or waking?},
  author={Benington, Joel H and Heller, H Craig},
  journal={Prog Neurobiol},
  volume={44},
  pages={433--449},
  year={1994},
}

@article{benington1995apamin,
  title={Apamin, a selective {SK} potassium channel blocker, suppresses {REM} sleep without a compensatory rebound},
  author={Benington, Joel H and Woudenberg, M Catherine and Heller, H Craig},
  journal={Brain Research},
  volume={692},
  pages={86--92},
  year={1995}
}

@article{barbato1998homeostatic,
  title={Homeostatic regulation of {REM} sleep in humans during extended sleep},
  author={Barbato, Giuseppe and Wehr, Thomas A},
  journal={Sleep},
  volume={21},
  number={3},
  pages={267--276},
  year={1998},
  publisher={Oxford University Press}
}

@article{cajochen2024ultradian,
  title={Ultradian sleep cycles: Frequency, duration, and associations with individual and environmental factors—A retrospective study},
  author={Cajochen, Christian and Reichert, Carolin Franziska and M{\"u}nch, Mirjam and Gabel, Virginie and Stefani, Oliver and Chellappa, Sarah Laxhmi and Schmidt, Christina},
  journal={Sleep Health},
  volume={10},
  number={1},
  pages={S52--S62},
  year={2024},
  publisher={Elsevier}
}

@incollection{heller2021regulation,
  title={The regulation of sleep},
  author={Heller, Craig},
  booktitle={Oxford Research Encyclopedia of Neuroscience},
  year={2021},
  publisher= {Oxford University Press}
}

@article{dijk2010age,
  title={Age-related reduction in daytime sleep propensity and nocturnal slow wave sleep},
  author={Dijk, Derk-Jan and Groeger, John A and Stanley, Neil and Deacon, Stephen},
  journal={Sleep},
  volume={33},
  number={2},
  pages={211--223},
  year={2010},
  publisher={Oxford University Press}
}

@article{ursin1970sleep,
  title={Sleep stage relations within the sleep cycles of the cat},
  author={Ursin, Reidun},
  journal={Brain research},
  volume={20},
  number={1},
  pages={91--97},
  year={1970},
  publisher={Elsevier}
}

@article{merica1991study,
  title={A study of the interrupted {REM} episode},
  author={Merica, H and Gaillard, J-M},
  journal={Physiology \& behavior},
  volume={50},
  number={6},
  pages={1153--1159},
  year={1991},
  publisher={Elsevier}
}

@article{kripke1968nocturnal,
  title={Nocturnal sleep in rhesus monkeys},
  author={Kripke, DF and Reite, ML and Pegram, GV and Stephens, LM and Lewis, OF},
  journal={Electroencephalography and Clinical Neurophysiology},
  volume={24},
  number={6},
  pages={581--586},
  year={1968},
  publisher={Elsevier}
}

@article{chang2015evening,
  title={Evening use of light-emitting eReaders negatively affects sleep, circadian timing, and next-morning alertness},
  author={Chang, Anne-Marie and Aeschbach, Daniel and Duffy, Jeanne F and Czeisler, Charles A},
  journal={Proceedings of the National Academy of Sciences},
  volume={112},
  number={4},
  pages={1232--1237},
  year={2015},
  publisher={National Academy of Sciences}
}

@article{nielsen2010rem,
  title={{REM} sleep characteristics of nightmare sufferers before and after {REM} sleep deprivation},
  author={Nielsen, Tore A and Paquette, Tyna and Solomonova, Elizaveta and Lara-Carrasco, Jessica and Popova, Ani and Levrier, Katia},
  journal={Sleep Medicine},
  volume={11},
  number={2},
  pages={172--179},
  year={2010},
  publisher={Elsevier}
}

@article{bassi2009time,
  title={The time course of the probability of transition into and out of {REM} sleep},
  author={Bassi Acu{\~n}a, Alejandro and Vivaldi V{\'e}jar, Ennio and Ocampo Garc{\'e}s, Adri{\'a}n},
  journal={Sleep},
  year={2009},
  publisher={American Academy of Sleep Medicine}
}

@article{amici1994pattern,
  title={Pattern of desynchronized sleep during deprivation and recovery induced in the rat by changes in ambient temperature},
  author={Amici, Roberto and Zamboni, Giovanni and Perez, Emanuele and Jones, Christine A and Toni, IVAN and Culin, Fabio and Parmeggiani, Pier Luigi},
  journal={Journal of sleep research},
  volume={3},
  number={4},
  pages={250--256},
  year={1994},
  publisher={Wiley Online Library}
}

@article{gregory2002two,
  title={A two-state stochastic model of {REM} sleep architecture in the rat},
  author={Gregory, Gavin G and Cabeza, Rafael},
  journal={Journal of neurophysiology},
  volume={88},
  number={5},
  pages={2589--2597},
  year={2002},
  publisher={American Physiological Society Bethesda, MD}
}

@article{sleep_edf_kemp,
  title={Analysis of a sleep-dependent neuronal feedback loop: the slow-wave microcontinuity of the EEG},
  author={Kemp, B and Zwinderman, A H and Tuk, B and Kamphuisen, H A C and Oberye, J J L},
  journal={IEEE-BME},
  volume={47},
  number={9},
  pages={1185-1194},
  year={2000}
}

@article{sleep_edf_goldberger,
  title={PhysioBank, PhysioToolkit, and PhysioNet: Components of a new research resource for complex physiologic signals},
  author={Goldberger, A and Amaral, L and Glass, L and Hausdorff, J and Ivanov, P C and Mark, R and Mietus, J E and Moody, G B and Peng, C K and Stanley, H E},
  journal={Circulation [Online]},
  volume={101},
  number={23},
  pages={e215-e220},
  year={2000}
}

@article{mnc_zhang,
  title={The National Sleep Research Resource: towards a sleep data commons},
  author={Zhang, G Q and Cui, L and Mueller, R and Tao, S and Kim, M and Rueschman, M and Mariani, S and Mobley, D and Redline, S},
  journal={J Am Med Inform Assoc},
  volume={25},
  number={10},
  pages={1351-1358},
  year={2018}
}

@article{mnc_stephansen,
  title={Neural network analysis of sleep stages enables efficient diagnosis of narcolepsy},
  author={Stephansen, J B and Olesen, A N and Olsen, M and Ambati, A and Leary, E B and Moore, H E and Carrillo, O and Lin, L and Han, F and Yan, H and Sun, Y L and Dauvilliers, Y and Scholz, S and Barateau, L and Hogl, B and Stefani, A and Hong, S C and Kim, T W and Pizza, F and Plazzi, G and Vandi, S and Antelmi, E and Perrin, D and Kuna, S T and Schweitzer, P K and Kushida, C and Peppard, P E and Sorensen, H B D and Jennum, P and Mignot, E},
  journal={Nat Commun},
  volume={9},
  number={1},
  pages={5229},
  year={2018}
}

@misc{bitbrain_dataset,
  author = {Eduardo López-Larraz AND María Sierra-Torralba AND Sergio Clemente AND Galit Fierro AND David Oriol AND Javier Minguez AND Luis Montesano AND Jens G. Klinzing},
  title = {Bitbrain Open Access Sleep Dataset},
  year = {2025},
  doi = {doi:10.18112/openneuro.ds005555.v1.1.0},
  publisher = {OpenNeuro}
}

@article{silverstein_ratdata,
title = {Effect of prolonged sedation with dexmedetomidine, midazolam, propofol, and sevoflurane on sleep homeostasis in rats},
journal = {British Journal of Anaesthesia},
volume = {132},
number = {6},
pages = {1248-1259},
year = {2024},
author = {Brian H. Silverstein and Anjum Parkar and Trent Groenhout and Zuzanna Fracz and Anna M. Fryzel and Christopher W. Fields and Amanda Nelson and Tiecheng Liu and Giancarlo Vanini and George A. Mashour and Dinesh Pal}
}

@article{esposito2004,
title = {Single and Sequential {REM} sleep episodes in humans: a phylogenetic left-over?},
journal = {Neurosci Lett},
volume = {368},
number = {1},
pages = {52–56},
year = {2004},
author = {Esposito, M. J. and Zamboni, G. and Natale, V. and Lucidi, F. and Devoto, A. and Violani, C.}
}

@article{esposito2003,
title = {Inter-{REM} sleep intervals distribution in healthy young subjects},
journal = {Sleep and Hypnosis},
volume = {5},
number = {1},
pages = {1-6},
year = {2003},
author = {Esposito, M. J. and Natale, V. and Occhionero, M. and Cicogna, P.}
}

@article{LUPPI2024101907,
title = {Which structure generates paradoxical ({REM}) sleep: The brainstem, the hypothalamus, the amygdala or the cortex?},
journal = {Sleep Medicine Reviews},
volume = {74},
pages = {101907},
year = {2024},
issn = {1087-0792},
doi = {https://doi.org/10.1016/j.smrv.2024.101907},
url = {https://www.sciencedirect.com/science/article/pii/S108707922400011X},
author = {Pierre-Hervé Luppi and Amarine Chancel and Justin Malcey and Sébastien Cabrera and Patrice Fort and Renato M. Maciel}
}

@article{kobayashi1985sleep,
  title={Sleep cycle as a basic unit of sleep},
  author={Kobayashi, T},
  journal={Ultradian rhythms in physiology and behavior},
  pages={261--269},
  year={1985},
  publisher={Springer-Verlag}
}

@article{stute1993bootstrap,
  title={Bootstrap based goodness-of-fit-tests},
  author={Stute, Winfried and Manteiga, Wenceslao Gonz{\'a}les and Quindimil, Manuel Presedo},
  journal={Metrika},
  volume={40},
  number={1},
  pages={243--256},
  year={1993},
  publisher={Springer}
}

@article{charles1980glossary,
  title={Glossary of standardized terminology for sleepbiological rhythm research},
  author={Charles, A Czeisler},
  journal={Sleep},
  volume={2},
  number={3},
  pages={287--288},
  year={1980},
  publisher={Oxford University Press}
}

@article{stephansen2018neural,
  title={Neural network analysis of sleep stages enables efficient diagnosis of narcolepsy},
  author={Stephansen, Jens B and Olesen, Alexander N and Olsen, Mads and Ambati, Aditya and Leary, Eileen B and Moore, Hyatt E and Carrillo, Oscar and Lin, Ling and Han, Fang and Yan, Han and others},
  journal={Nature communications},
  volume={9},
  number={1},
  pages={5229},
  year={2018},
  publisher={Nature Publishing Group UK London}
}

@article{christensen2017novel,
  title={Novel method for evaluation of eye movements in patients with narcolepsy},
  author={Christensen, Julie AE and Kempfner, Lykke and Leonthin, Helle L and Hvidtfelt, Mathias and Nikolic, Miki and Kornum, Birgitte Rahbek and Jennum, Poul},
  journal={Sleep Medicine},
  volume={33},
  pages={171--180},
  year={2017},
  publisher={Elsevier}
}

@techreport{rechtschaffen1968manual,
    author = {A. Rechtschaffen, A. Kales},
    title = {A Manual of Standardized Terminology, Techniques and Scoring System of Sleep Stages in Human Subjects},
    institution = {(Brain Information Service/Brain 
639 Research Institute, University of California},
    year = {1968}
}

@misc{berry2015aasm,
  title={AASM scoring manual version 2.2 updates: new chapters for scoring infant sleep staging and home sleep apnea testing},
  author={Berry, Richard B and Gamaldo, Charlene E and Harding, Susan M and Brooks, Rita and Lloyd, Robin M and Vaughn, Bradley V and Marcus, Carole L},
  journal={Journal of Clinical Sleep Medicine},
  volume={11},
  number={11},
  pages={1253--1254},
  year={2015},
  publisher={American Academy of Sleep Medicine}
}

@article{andlauer2013nocturnal,
  title={Nocturnal rapid eye movement sleep latency for identifying patients with narcolepsy/hypocretin deficiency},
  author={Andlauer, Olivier and Moore IV, Hyatt and Jouhier, Laura and Drake, Christopher and Peppard, Paul E and Han, Fang and Hong, Seung-Chul and Poli, Francesca and Plazzi, Giuseppe and O’Hara, Ruth and others},
  journal={JAMA neurology},
  volume={70},
  number={7},
  year={2013}
}

@article{moore2014design,
  title={Design and validation of a periodic leg movement detector},
  author={Moore IV, Hyatt and Leary, Eileen and Lee, Seo-Young and Carrillo, Oscar and Stubbs, Robin and Peppard, Paul and Young, Terry and Widrow, Bernard and Mignot, Emmanuel},
  journal={PloS one},
  volume={9},
  number={12},
  pages={e114565},
  year={2014},
  publisher={Public Library of Science San Francisco, USA}
}


\newpage
\section*{Supplementary Materials}\label{sec: supp}

\setcounter{section}{0}
\renewcommand{\thesection}{S\arabic{section}}

\section{Mouse and Rat (Dark phase) Results}\label{dark phase}

\renewcommand\thefigure{S\arabic{figure}}\setcounter{figure}{0}
\begin{figure}[ht!]
    \centering
    \includegraphics[width=1\linewidth]{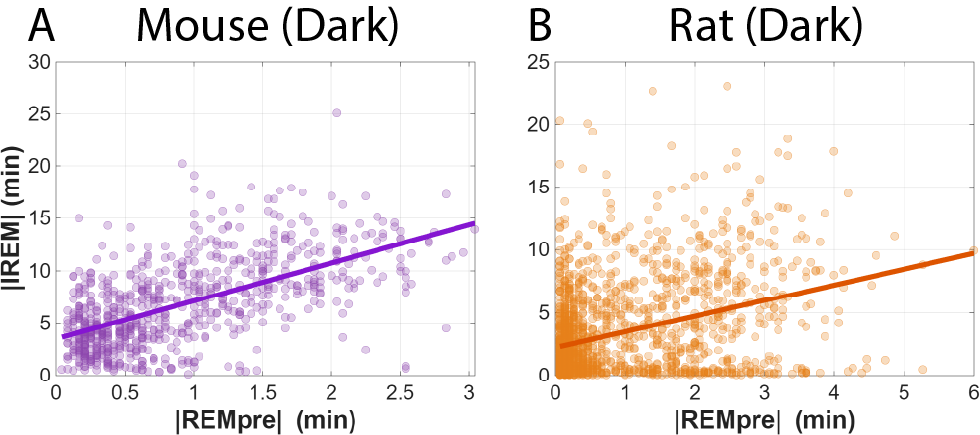}
    \caption{{\bf Inter-REMS interval duration \irem \ versus preceding REMS bout duration \rempre \ in mice and rats (dark phase).}
Scatter plots show the relationship between the duration of the preceding REMS bout (\rempre) and the length of the subsequent inter-REMS interval (\irem) \ for (A) mouse  and (B) rat data in the dark phase. 
Each point represents a single REMS cycle. Both species exhibit a positive association, indicated by the fitted regression line, demonstrating that longer inter-REMS intervals tend to follow longer \rempre \ bouts (Mouse Dark: $p\approx 1.8\times10^{-77}$ and Rat Dark: $p\approx 8.44\times 10^{-55}$).}
    \label{fig: S1_dark}
\end{figure}


\begin{figure}[htp!]
    \centering
    \includegraphics[width=1\linewidth]{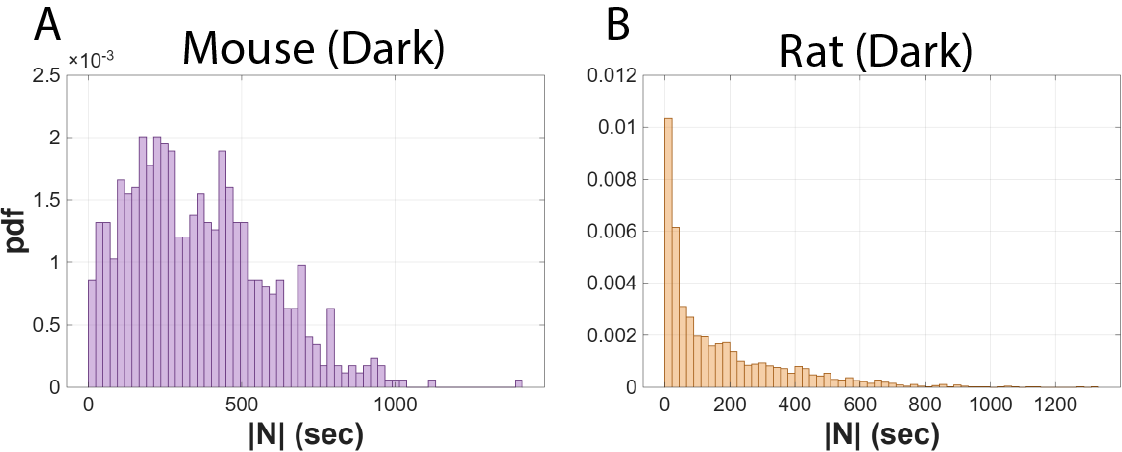}
    \caption{{\bf{Empirical cumulative inter-REMS NREMS duration $|N|$ distributions in mice and rats (dark phase).}}
Histograms of $|N|$ values pooled across all REMS cycles for (A) mouse and (B) rat data in the dark phase.  }
    \label{fig: S2_dark}
\end{figure}


\begin{figure}
    \centering
    \includegraphics[width=1\linewidth]{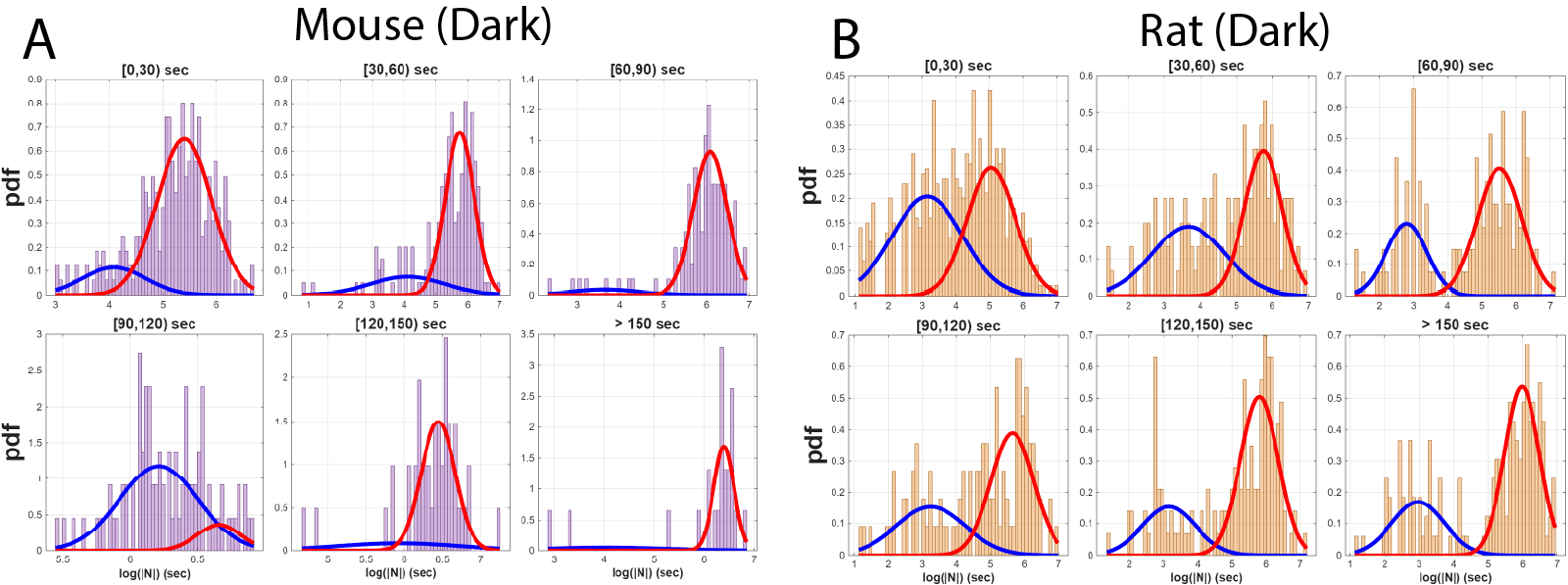}
\caption{{\bf{Gaussian mixture modeling (GMM) fits of the inter-REMS NREMS durations $|N|$ binned by \rempre \ duration for rodent data during the dark phase.}}
(A) mouse (dark) and (B) rat (dark). Each panel shows the distribution of $\log(|N|)$ within a specific \rempre \ bin (labeled above each subplot). Within every bin, a two-component GMM is fit to the $\log(|N|)$ durations: the short-interval (sequential) mode (blue curve) and the long-interval (single) mode (red curve). 
}%
\label{fig: S3_dark}%
\end{figure}


\newpage

\setcounter{table}{0}
\renewcommand{\thetable}{S\arabic{table}}
\begin{table}[ht!]
\centering
\caption{Two-component Gaussian mixture model (GMM) parameters for mouse light-phase data after long-wake filtering ($\geq 2$ min). The model was fit to $z=\log(|N|)$, where $|N|$ is NREMS duration in seconds. For each REM${\mathrm{pre}}$ bin, $n$ is the number of REM cycles included, $w_j$ is the mixing weight, $\mu_j$ is the mean, and $\sigma_j$ is the standard deviation of component $j$ ($j=1,2$) on the log scale. Components are ordered by increasing mean.}
\label{tab:gmm_mouse_light}
\begin{tabular}{c c c c c c c c c}
\hline
Bin & REM${\mathrm{pre}}$ range (s) & $n$ & $w_1$ & $\mu_1$ & $\sigma_1$ & $w_2$ & $\mu_2$ & $\sigma_2$ \\
\hline
1 & $[0,30)$     & 1764 & 0.570364 & 4.679818 & 0.747388 & 0.429636 & 5.584552 & 0.428502 \\
2 & $[30,60)$    & 955  & 0.274363 & 4.139118 & 0.795511 & 0.725637 & 5.699581 & 0.457006 \\
3 & $[60,90)$    & 538  & 0.211115 & 3.962436 & 1.006828 & 0.788885 & 6.073470 & 0.374800 \\
4 & $[90,120)$   & 359  & 0.081130 & 3.715733 & 0.997149 & 0.918870 & 6.241783 & 0.335687 \\
5 & $[120,150)$  & 214  & 0.060897 & 3.497884 & 0.701369 & 0.939103 & 6.341828 & 0.307352 \\
6 & $[150,\infty)$ & 175 & 0.060199 & 5.989144 & 0.585698 & 0.939801 & 6.501553 & 0.254413 \\
\hline
\end{tabular}
\end{table}

\begin{table}[ht!]
\centering
\caption{Two-component Gaussian mixture model (GMM) parameters for mouse dark-phase data after long-wake filtering ($\geq 2$ min). The model was fit to $z=\log(|N|)$, where $|N|$ is NREMS duration in seconds. For each REM${\mathrm{pre}}$ bin, $n$ is the number of REM cycles included, $w_j$ is the mixing weight, $\mu_j$ is the mean, and $\sigma_j$ is the standard deviation of component $j$ ($j=1,2$) on the log scale. Components are ordered by increasing mean.}
\label{tab:gmm_mouse_dark}
\begin{tabular}{c c c c c c c c c}
\hline
Bin & REM${\mathrm{pre}}$ range (s) & $n$ & $w_1$ & $\mu_1$ & $\sigma_1$ & $w_2$ & $\mu_2$ & $\sigma_2$ \\
\hline
1 & $[0,30)$      & 263 & 0.175742 & 4.118561 & 0.609007 & 0.824258 & 5.405572 & 0.508734 \\
2 & $[30,60)$     & 190 & 0.222775 & 4.107440 & 1.201124 & 0.777225 & 5.755275 & 0.459099 \\
3 & $[60,90)$     & 126 & 0.072323 & 3.638997 & 0.809157 & 0.927677 & 6.093499 & 0.397896 \\
4 & $[90,120)$    & 90  & 0.854246 & 6.212131 & 0.291010 & 0.145754 & 6.656193 & 0.159850 \\
5 & $[120,150)$   & 47  & 0.168139 & 5.926227 & 0.816934 & 0.831861 & 6.444326 & 0.221955 \\
6 & $[150,\infty)$ & 23 & 0.142958 & 4.037990 & 1.207258 & 0.857042 & 6.400520 & 0.202394 \\
\hline
\end{tabular}
\end{table}

\begin{table}[ht!]
\centering
\caption{Two-component Gaussian mixture model (GMM) parameters for rat light-phase data after long-wake filtering ($\geq 2$ min). The model was fit to $z=\log(|N|)$, where $|N|$ is NREMS duration in seconds. For each REM${\mathrm{pre}}$ bin, $n$ is the number of REM cycles included, $w_j$ is the mixing weight, $\mu_j$ is the mean, and $\sigma_j$ is the standard deviation of component $j$ ($j=1,2$) on the log scale. Components are ordered by increasing mean.}
\label{tab:gmm_rat_light}
\begin{tabular}{c c c c c c c c c}
\hline
Bin & REM${\mathrm{pre}}$ range (s) & $n$ & $w_1$ & $\mu_1$ & $\sigma_1$ & $w_2$ & $\mu_2$ & $\sigma_2$ \\
\hline
1 & $[0,30)$      & 1218 & 0.564796 & 3.130091 & 1.064274 & 0.435204 & 5.091941 & 0.687526 \\
2 & $[30,60)$     & 217  & 0.504750 & 3.770096 & 1.129371 & 0.495250 & 5.716675 & 0.526266 \\
3 & $[60,90)$     & 152  & 0.357612 & 2.829082 & 0.682314 & 0.642388 & 5.501217 & 0.693970 \\
4 & $[90,120)$    & 129  & 0.545531 & 3.692626 & 1.203597 & 0.454469 & 5.878608 & 0.426530 \\
5 & $[120,150)$   & 171  & 0.348783 & 3.318593 & 0.925519 & 0.651217 & 5.866989 & 0.504289 \\
6 & $[150,\infty)$ & 202 & 0.325012 & 2.920484 & 0.809527 & 0.674988 & 6.014627 & 0.493327 \\
\hline
\end{tabular}
\end{table}

\begin{table}[ht!]
\centering
\caption{Two-component Gaussian mixture model (GMM) parameters for rat dark-phase data after long-wake filtering ($\geq 2$ min). The model was fit to $z=\log(|N|)$, where $|N|$ is NREMS duration in seconds. For each REM${\mathrm{pre}}$ bin, $n$ is the number of REM cycles included, $w_j$ is the mixing weight, $\mu_j$ is the mean, and $\sigma_j$ is the standard deviation of component $j$ ($j=1,2$) on the log scale. Components are ordered by increasing mean.}
\label{tab:gmm_rat_dark}
\begin{tabular}{c c c c c c c c c}
\hline
Bin & REM${\mathrm{pre}}$ range (s) & $n$ & $w_1$ & $\mu_1$ & $\sigma_1$ & $w_2$ & $\mu_2$ & $\sigma_2$ \\
\hline
1 & $[0,30)$      & 1012 & 0.536597 & 3.163676 & 1.059356 & 0.463403 & 5.046773 & 0.712614 \\
2 & $[30,60)$     & 162  & 0.481607 & 3.678129 & 1.007557 & 0.518393 & 5.755845 & 0.524027 \\
3 & $[60,90)$     & 140  & 0.338321 & 2.814225 & 0.554891 & 0.661679 & 5.515701 & 0.651625 \\
4 & $[90,120)$    & 115  & 0.373850 & 3.174141 & 0.984049 & 0.626150 & 5.615120 & 0.638786 \\
5 & $[120,150)$   & 147  & 0.301097 & 3.174769 & 0.789621 & 0.698903 & 5.814754 & 0.550853 \\
6 & $[150,\infty)$ & 169 & 0.321076 & 2.974512 & 0.744634 & 0.678924 & 5.986087 & 0.504694 \\
\hline
\end{tabular}
\end{table}



\begin{figure}[H]
    \centering
    \includegraphics[width=1\linewidth]{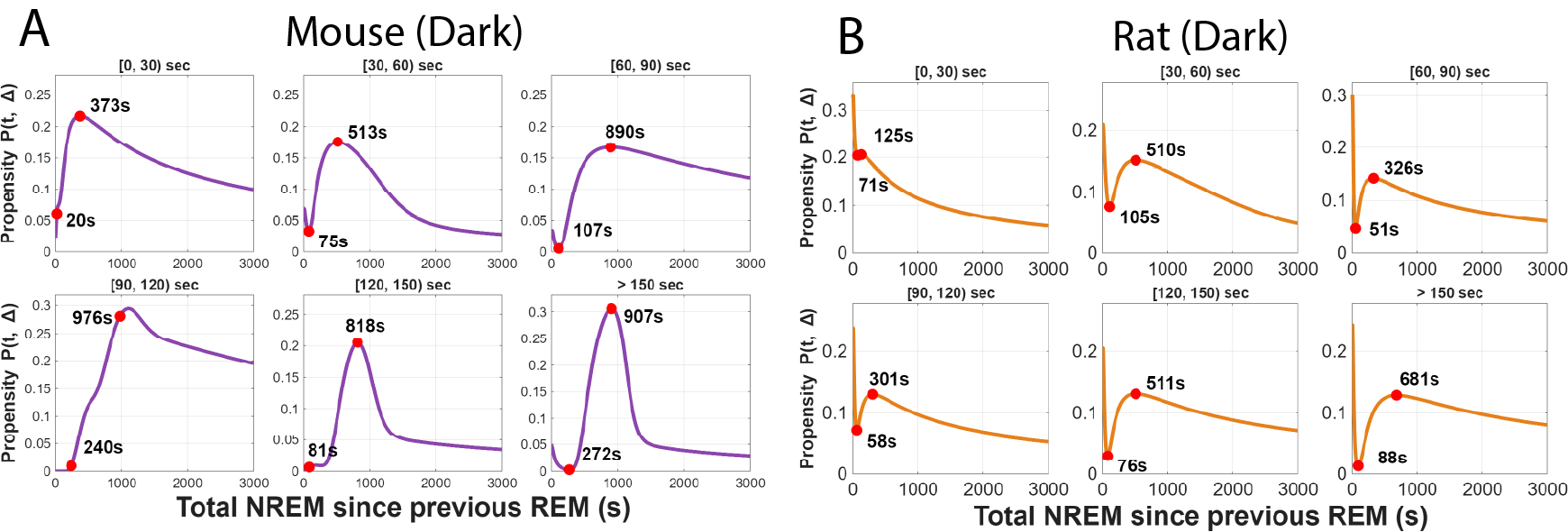}
\caption{{\bf REMS propensity across \rempre\ bins for (A) mouse and (B) rat in the dark phase.}
For each species, the propensity function $P(t,\Delta)$ is computed within \rempre-stratified bins.}
\label{fig: S5_dark}%
\end{figure}


\begin{figure}[H]
    \centering
    \includegraphics[width=1\linewidth]{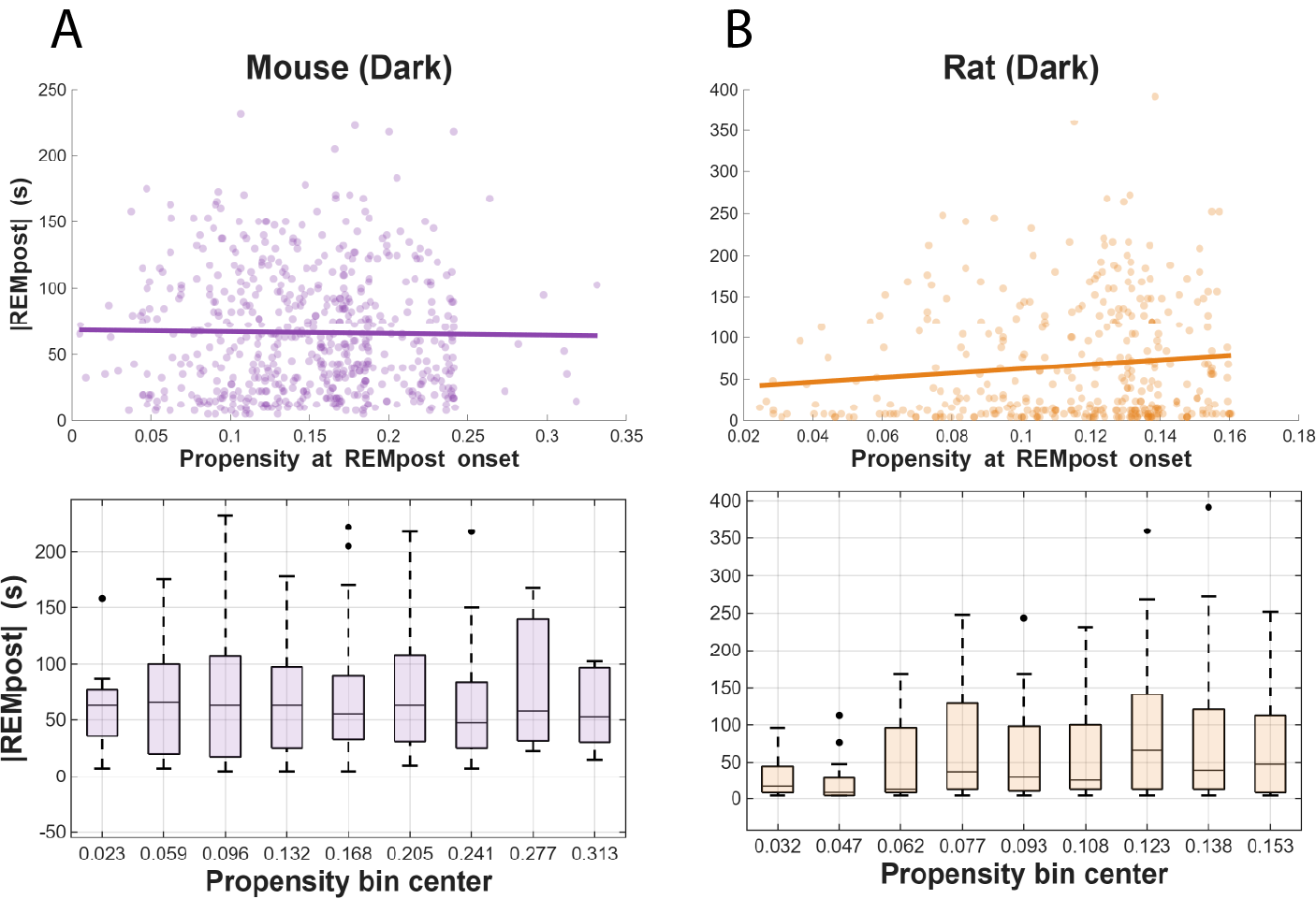}
\caption{{\bf Top: Correlation between REMS propensity and the duration of the subsequent REMS bout \rempost\ for mice and rats dark-phase.}
Scatter plots show the relationship between the REMS propensity at REMS onset and the duration of the following REMS episode ($|\mathrm{REM{post}}|$) (the p-values $p=0.69$ (mouse dark) and $p=0.02$ (rat dark). \textbf{Bottom: REM${\mathrm{post}}$ duration vs.\ REMS propensity at onset.} Box-and-whisker summaries of $|\mathrm{REM}{\mathrm{post}}|$ (s) across propensity bins.
}
  \label{fig: S6_dark}
\end{figure}

\section{Definition and Filtering of Long-Wake Episodes}\label{sec: supp_filter}
During sleep recordings, rodents and human periodically exhibit spontaneous wake periods between consecutive NREM–REM cycles. These wake bouts vary in duration, and exceptionally long episodes can perturb the temporal rhythm of the sleep cycle. To quantify their effect, we introduced a \textbf{Long-Wake (LW) filtering procedure} in which cycles containing a contiguous wake bout greater than a given threshold were excluded from subsequent analyses. The goal of this filtering step is to identify which threshold yields the most stable, well-structured distribution of NREM durations, \emph{i.e.}, which ``Long-Wake" cutoff best captures the intrinsic rhythm of sleep without distortions from abnormally long awakenings. Specifically, thresholds of $\tau_\text{thr} = \{2,\,5,\,7,\,10\}$ minutes were applied to human and both {mouse} and {rat} datasets under light and dark conditions. For each case, cycles satisfying
``max contiguous wake duration $\geq \tau_\text{thr}$", 
were discarded, and all remaining cycles were retained for Gaussian Mixture Model (GMM) fitting of NREM durations.

\subsection{Assessing Goodness of Fit: KS Test vs Model Selection Criteria}
To evaluate how well the Gaussian Mixture Model (GMM) describes the observed NREM duration data, one commonly used nonparametric measure is the {Kolmogorov-Smirnov (KS) statistic}. The KS statistic quantifies the largest vertical distance between the empirical cumulative distribution function (ECDF) of the data and the theoretical cumulative distribution function (CDF) predicted by the fitted model:
\[
D_{\mathrm{KS}} = \sup_x \, \big| F_\text{emp}(x) - F_\text{model}(x) \big|,
\]
where $F_\text{emp}(x)$ is the empirical CDF of the sample and $F_\text{model}(x)$ is the CDF derived from the GMM.  We use the KS statistic to carry out a KS test corrected for the GMM, similarly to how the Lilliefors-corrected KS test corrects for a Gaussian distribution. In the corrected KS-test we use Monte-Carlo simulations to compute the probability distribution of 
\[
\tilde{D}_{\mathrm{KS}} = \sup_x \, \big| F_\text{test}(x) - F_\text{model}(x) \big|,
\]
where $F_\text{test}(x)$ is an empirical CDF produced from $n$ measurements of a random variable with CDF given by $F_\text{model}(x)$, $n$ being equal to the number of samples used to compute $F_\text{emp}(x)$. If 
\[
\text{Probability}\left(\tilde{D}_{\mathrm{KS}} > D_{\mathrm{KS}} \right) < 0.05,
\]
then the corrected KS-test rejects the null hypothesis that the fitted model is the distribution that produced the empirical CDF. In this context, 
\[
\text{p-value} = \text{Probability}\left(\tilde{D}_{\mathrm{KS}} > D_{\mathrm{KS}} \right). 
\]
Intuitively, the larger the $D_{\mathrm{KS}}$, the worse the fit, and the smaller the p-value. Because smaller p-values for the corrected KS test indicate a worse fit, their interpretation differs from most goodness-of-fit tests, where lower values are desired. 

Although the KS statistic is conceptually simple and distribution free, it is mainly sensitive to differences in the \emph{central region} of the CDF, while being less responsive to discrepancies in the tails.  
Furthermore, when two parametric models are both close approximations to the same empirical distribution (as in our GMMs under different Long-Wake thresholds), their $D_{\mathrm{KS}}$ values differ only marginally, often within the range of sampling noise (See tables~\ref{tab:ks_distances_ksbest_bins_by_species}-~\ref{tab:human_all}).
Therefore, to obtain a more discriminative criterion that accounts for both fit accuracy and model complexity, we also used penalized-likelihood measures such as the {``Bayesian Information Criterion (BIC)"}, which better reflect the trade-off between descriptive power and parsimony in parametric models.


The KS-best distances in Table~\ref{tab:ks_distances_ksbest_bins_by_species} quantify the maximum discrepancy between the empirical CDF of $z=\log(|N|)$ (log-NREM duration) and the fitted two-component GMM CDF, after selecting the Expectation-Maximization (EM) algorithm restart that minimizes this discrepancy. Across datasets, the lumped fits have small KS distances ($D \approx 0.0187$–$0.0449$), indicating that a 2-component GMM provides a close overall approximation to the distribution of log-NREM durations under the 2-min long-wake filtering rule. Within REMpre bins, most KS distances remain modest (typically $D\lesssim 0.06$), suggesting that the mixture model captures bin-specific distributions reasonably well. The largest discrepancies occur in later REMpre bins for darker-phase mouse data (Bin 6: $D=0.0908$) and in Rat (Dark) Bin 4 ($D=0.0884$), consistent with reduced sample sizes and/or more heterogeneous dynamics in the tail bins; nevertheless, these values still represent at most a $\sim$9\% maximum CDF deviation. Overall, the table supports the conclusion that the fitted 2-GMMs track the empirical distributions of $\log(|N|)$ well across species/phase conditions, with the primary departures concentrated in a small number of tail bins.

\begin{table}[ht]
\centering
\caption{Kolmogorov--Smirnov (KS) distances $D$ between empirical distributions of $z=\log(|N|)$ and fitted two--component Gaussian mixture models (2--GMM). Long-wake filtering: $\ge 2$ min. Columns are species/phase; rows are REM$_{pre}$ bins (plus lumped).}
\label{tab:ks_distances_ksbest_bins_by_species}
\small
\begin{tabular}{lcccc}
\hline
 & Mouse (Light) & Mouse (Dark) & Rat (Light) & Rat (Dark) \\
\hline
Lumped $D$ & 0.018669 & 0.044872 & 0.021688 & 0.020322 \\
\hline
Bin 1 $D$ & 0.012145 & 0.022943 & 0.032828 & 0.028894 \\
Bin 2 $D$ & 0.029810 & 0.056980 & 0.022256 & 0.028700 \\
Bin 3 $D$ & 0.037001 & 0.035763 & 0.035950 & 0.037927 \\
Bin 4 $D$ & 0.027220 & 0.053244 & 0.053032 & 0.088368 \\
Bin 5 $D$ & 0.030804 & 0.067448 & 0.056791 & 0.069921 \\
Bin 6 $D$ & 0.047724 & 0.090825 & 0.040919 & 0.038563 \\
\hline
\end{tabular}
\end{table}
All four rodent datasets show KS diagnostic $p$-values exceeding $0.05$ for the lumped fits and for every REMpre bin (Table~\ref{tab:ks_pvalues_bins_by_species}), indicating no bin exhibits an obvious distributional mismatch between the empirical CDF of $z=\log(|N|)$ and the fitted two-component GMM. At the pooled (lumped) level, the $p$-values are moderate (Mouse Light: $0.11$–$0.12$; Mouse Dark: $\approx 0.10$; Rat Light: $0.28$; Rat Dark: $0.46$). Within bins, most $p$-values are large (often $>0.3$ and frequently near $1$), suggesting the fitted mixture CDF tracks the empirical CDF closely; the smallest bin-level $p$-values occur in Rat Light bin~1 ($p=0.142$) and Rat Dark bin~4 ($p=0.316$), but remain above the $0.05$ threshold. Overall, these KS diagnostics support the adequacy of the two-component GMM approximation to log-NREM durations across REMpre bins under the 2-min long-wake filtering rule (noting that KS $p$-values are used here as descriptive goodness-of-fit checks when model parameters are estimated from the data).

\begin{table}[ht]
\centering
\caption{Kolmogorov--Smirnov (KS) $p$-values for two-component GMM fits to $z=\log(|N|)$ (NREM duration in seconds), after long-wake filtering ($\ge 2$ min). Columns are species/phase; rows are REMpre bins (plus lumped).}
\label{tab:ks_pvalues_bins_by_species}
\small
\begin{tabular}{l|cccc}
\hline
 & Mouse (Light) & Mouse (Dark) & Rat (Light) & Rat (Dark) \\
\hline
Lumped $p$ & 0.1213 & 0.0993 & 0.2766 & 0.4632 \\
\hline
Bin 1 $p$ & 0.9560 & 0.9990 & 0.1421 & 0.3621 \\
Bin 2 $p$ & 0.3590 & 0.5557 & 0.9999 & 0.9992 \\
Bin 3 $p$ & 0.4458 & 0.9965 & 0.9881 & 0.9862 \\
Bin 4 $p$ & 0.9503 & 0.9557 & 0.8519 & 0.3164 \\
Bin 5 $p$ & 0.9859 & 0.9793 & 0.6273 & 0.4551 \\
Bin 6 $p$ & 0.8112 & 0.9879 & 0.8813 & 0.9598 \\
\hline
\end{tabular}
\end{table}

\subsection{Bayesian Information Criterion (BIC) Formulation}
For each dataset, both a {lumped} model (one global 2-component GMM) and a {per–bin} model (2-component GMM fitted separately within each REM preceding bin) were evaluated.  


The Bayesian Information Criterion (BIC) is defined as
\[
\mathrm{BIC} = -2 \ln(\hat{L}) + p \ln(N),
\]
where $\hat{L}$ is the maximized likelihood of the fitted model, $p$ is the number of free parameters, and $N$ is the number of data points (that is, the number of cycles used in fitting). Lower BIC values indicate a better trade-off between goodness of fit and model complexity.

For a one-dimensional Gaussian mixture model with $K=2$ components, the number of free parameters is
\[
p = (K-1) + 2K.
\]
Here, $(K-1)$ counts the independent mixing proportions, since the $K$ component weights must sum to 1, leaving only $K-1$ free. The term $2K$ counts the Gaussian parameters: each of the $K$ components contributes one mean and one variance. Thus, for a two-component model,
\[
p = (2-1) + 2(2) = 1 + 4 = 5.
\]

For the per-bin model with six REM${}_{\mathrm{pre}}$ bins, a separate two-component GMM is fit in each bin, so the total number of free parameters is
\[
p_{\text{per-bin}} = 6 \times 5 = 30.
\]

\subsection{BIC Results Across Long-Wake Thresholds}
Tables~\ref{tab:mouse_light}-~\ref{tab:human_all} summarize the BIC and KS results for mouse and rat datasets under Light and Dark conditions and human (note LL indicates Likelihood, BIC is Bayesian Information Criterion, and KS is Kolmogrov-Smirnov test).

\begin{table}[H]
\centering
\caption{Mouse (Light) dataset under different long–Wake thresholds.}
\label{tab:mouse_light}
\begin{tabular}{ccrrrr}
\hline
\textbf{Thr (min)} & \textbf{Model} & \textbf{N} & \textbf{LL} & \textbf{BIC} & \textbf{KS} \\
\hline
2  & Lumped  & 4497 & -5582.6 & 11207.0 & 0.0259 \\
2  & Per-bin & 4497 & -4400.0 & {\textbf{9052.4}} & 0.0266 \\
5  & Lumped  & 4673 & -5824.3 & 11691.0 & 0.0250 \\
5  & Per-bin & 4673 & -4638.7 & 9530.8 & 0.0268 \\
7  & Lumped  & 4744 & -5934.0 & 11910.0 & 0.0242 \\
7  & Per-bin & 4744 & -4757.8 & 9769.5 & 0.0261 \\
10 & Lumped  & 4836 & -6076.5 & 12195.0 & 0.0238 \\
10 & Per-bin & 4836 & -4873.4 & 10001.0 & 0.0263 \\
\hline
\end{tabular}
\end{table}

\begin{table}[H]
\centering
\caption{Mouse (Dark) dataset under different long–Wake thresholds.}
\label{tab:mouse_dark}
\begin{tabular}{ccrrrr}
\hline
\textbf{Thr (min)} & \textbf{Model} & \textbf{N} & \textbf{LL} & \textbf{BIC} & \textbf{KS} \\
\hline
2  & Lumped  & 884 & -981.9  & 1997.7 & 0.0306 \\
2  & Per-bin & 884 & -739.2  & {\textbf{1681.9}} & 0.0443 \\
5  & Lumped  & 923 & -1013.2 & 2060.6 & 0.0282 \\
5  & Per-bin & 923 & -778.5  & 1761.9 & 0.0440 \\
7  & Lumped  & 943 & -1033.3 & 2100.9 & 0.0276 \\
7  & Per-bin & 943 & -807.0  & 1819.4 & 0.0435 \\
10 & Lumped  & 969 & -1057.7 & 2149.9 & 0.0271 \\
10 & Per-bin & 969 & -830.9  & 1868.1 & 0.0455 \\
\hline
\end{tabular}
\end{table}

\begin{table}[H]
\centering
\caption{Rat (Light) dataset under different long–Wake thresholds.}
\label{tab:rat_light}
\begin{tabular}{ccrrrr}
\hline
\textbf{Thr (min)} & \textbf{Model} & \textbf{N} & \textbf{LL} & \textbf{BIC} & \textbf{KS} \\
\hline
2  & Lumped  & 2297 & -3876.4 & 7791.4 & 0.0224 \\
2  & Per-bin & 2297 & -3637.4 & {\textbf{7506.9}} & 0.0394 \\
5  & Lumped  & 2448 & -4132.6 & 8304.2 & 0.0213 \\
5  & Per-bin & 2448 & -3869.3 & 7972.7 & 0.0384 \\
7  & Lumped  & 2481 & -4185.4 & 8409.8 & 0.0214 \\
7  & Per-bin & 2481 & -3918.4 & 8071.4 & 0.0372 \\
10 & Lumped  & 2520 & -4237.1 & 8513.4 & 0.0219 \\
10 & Per-bin & 2520 & -3972.1 & 8179.2 & 0.0372 \\
\hline
\end{tabular}
\end{table}

\begin{table}[H]
\centering
\caption{Rat (Dark) dataset under different long–Wake thresholds.}
\label{tab:rat_dark}
\begin{tabular}{ccrrrr}
\hline
\textbf{Thr (min)} & \textbf{Model} & \textbf{N} & \textbf{LL} & \textbf{BIC} & \textbf{KS} \\
\hline
2  & Lumped  & 1918 & -3203.5 & 6444.8 & 0.0261 \\
2  & Per-bin & 1918 & -2991.4 & { \textbf{6209.7}} & 0.0400 \\
5  & Lumped  & 2034 & -3401.2 & 6840.4 & 0.0252 \\
5  & Per-bin & 2034 & -3164.0 & 6556.6 & 0.0410 \\
7  & Lumped  & 2062 & -3445.2 & 6928.5 & 0.0266 \\
7  & Per-bin & 2062 & -3203.1 & 6635.1 & 0.0393 \\
10 & Lumped  & 2098 & -3492.8 & 7023.8 & 0.0273 \\
10 & Per-bin & 2098 & -3236.7 & 6702.8 & 0.0409 \\
\hline
\end{tabular}
\end{table}

\begin{table}[H]
\centering
\caption{Human dataset under different long–Wake thresholds.}
\label{tab:human_all}
\begin{tabular}{ccrrrr}
\hline
\textbf{Thr (min)} & \textbf{Model} & \textbf{N} & \textbf{LL} & \textbf{BIC} & \textbf{KS} \\
\hline
2  & Lumped  & 3397 & -5644.8 & 11330 & 0.1134 \\
2  & Per-bin & 3397 & -5529.2 & {\textbf{11262}} & 0.1062 \\
5  & Lumped  & 3786 & -6360.5 & 12762 & 0.1021 \\
5  & Per-bin & 3786 & -6256.1 & {{12718}} & 0.0943 \\
7  & Lumped  & 3877 & -6506.9 & 13055 & 0.1010 \\
7  & Per-bin & 3877 & -6413.4 & {{13033}} & 0.0970 \\
10 & Lumped  & 3958 & -6650.7 & 13343 & 0.0985 \\
10 & Per-bin & 3958 & -6556.5 & {13320} & 0.0949 \\
\hline
\end{tabular}
\end{table}

\noindent
Across all datasets, the \textbf{2-minute Long–Wake threshold consistently yields the lowest BIC}, while KS differences remain negligible.  
This result shows that shorter Long–Wake filtering ($\geq 2$ min) best isolates the stable NREM–REM rhythm and removes irregular cycles that degrade the overall mixture structure.

The comparative analyses across all rodent datasets show that simple KS-based tests are insufficient for discriminating subtle
changes in model performance under different Long–Wake thresholds. In contrast, the Bayesian Information Criterion (BIC)
provides a more sensitive and integrative measure by penalizing overfitting while rewarding explanatory accuracy. The BIC
consistently identified the 2-minute Long–Wake threshold as yielding the best model fit for both Mouse and Rat, under both
light and dark phases. This suggests that filtering out wake bouts longer than approximately 2 minutes produces a more
homogeneous and physiologically coherent dataset of NREM cycles.
From a biological perspective, these findings have implications for understanding the homeostatic regulation of NREM
sleep. Long continuous wake episodes appear to disrupt the stability and timing of subsequent NREM periods, leading to
greater variance and irregularity in cycle duration. This increase in variability, confirmed through variance-ratio analysis and
probability density histograms, indicates that NREM homeostasis becomes temporarily dysregulated after extended Wake.
Such episodes may reflect an overshoot or delayed recovery in the sleep–wake control system, possibly linked to altered
neuronal excitability or reduced synchronization in cortical slow-wave activity.

Therefore, the 2-minute cutoff represents not only a statistical optimum but also a physiologically meaningful boundary
between short wake bouts and disruptive wake episodes. By excluding these prolonged wake episodes, the resulting data
more faithfully capture the intrinsic rhythm of sleep–wake alternation, allowing for more accurate modeling of NREM
homeostatic dynamics and IREM stability.

\section{Human Model comparison, selection, and propensity analysis}\label{sec: supp_human_fit}
\subsection*{S3.1. Candidate model families for inter--REM NREM duration}

Let $T = |N|$ denote the human inter--REM NREM duration (minutes). 
The empirical distribution of $T$ exhibits strong right skewness, heavy tails,
and a pronounced spike at the measurement floor $x_{\min}=0.5$ minutes.
To determine an appropriate statistical model, we considered several
increasingly flexible parametric families.

\paragraph{S3.1.1. Two-component continuous mixtures.}
As an initial approach, we considered standard two-component mixtures
defined on $(0,\infty)$:

\begin{enumerate}[label=(\alph*), leftmargin=*]
\item Exponential + Lognormal (EXP+LN),
\item Weibull + Lognormal (WEI+LN),
\item Lognormal + Lognormal (LN+LN).
\end{enumerate}

These models assume
\[
f(t) = w f_S(t) + (1-w) f_L(t),
\qquad 0<w<1,
\]
where $f_S$ represents a short-duration component and $f_L$
a long-duration component.
The exponential model imposes memoryless short-bout dynamics.
The Weibull allows flexible short-bout hazard shapes.
The lognormal components allow heavy-tailed behavior.

\paragraph{S3.1.2. Three-component mixtures.}
To capture possible intermediate-duration structure,
we also evaluated three-component families such as:
\[
\text{WEI+LN+LN}, \quad
\text{LN+LN+LN}, \quad
\text{WEI+LN+LNP},
\]
where LNP denotes a lognormal--Pareto heavy-tail extension.

\paragraph{S3.1.3. Models with an explicit atom at $x_{\min}$.}
Inspection of the empirical distribution revealed a substantial spike
at $x_{\min}=0.5$ minutes. When purely continuous mixtures were fit to these data,
the short-duration component was forced to account for the excess probability mass
at the measurement floor. This often led to sensitive or inconsistent estimates of
the short-component parameters across candidate models and fitting runs, and could
also degrade the fit to the remaining continuous part of the distribution, including
the upper tail. To separate the point mass at the measurement floor from the
continuous behavior for $T>x_{\min}$, we therefore considered models with an explicit
\emph{atom} at $x_{\min}$; that is, a point mass is assigned to the event $T=x_{\min}$:
\[
\Pr(T=x_{\min}) = a,
\qquad
\Pr(T>x_{\min}) = 1-a,
\]
while the remaining probability mass is modeled by a continuous mixture on
$[x_{\min},\infty)$.



\subsection*{S3.2. Model selection criteria}


Selecting an appropriate parametric model for human inter--REM NREM durations
requires balancing three distinct objectives:
\emph{(i)} accurate representation of the empirical distribution,
\emph{(ii)} parsimony and interpretability,
and \emph{(iii)} statistical validity of goodness-of-fit inference under parameter estimation.
No single metric captures all three simultaneously.
For this reason, we adopted a layered evaluation framework combining
information-theoretic criteria and distributional diagnostics.

{First, we use likelihood-based criteria to compare models relative to one another; these likelihood-based measures are described in Section~S2.1 below.}
These criteria evaluate how well a model explains the observed data while penalizing
unnecessary complexity. However, good relative performance does not guarantee
that a model provides an adequate absolute description of the data.

Second, we compute global distributional discrepancy measures
that compare the fitted cumulative distribution function (CDF)
directly against the empirical CDF, {as described in S2.2 below}.
These metrics assess whether systematic deviations remain,
particularly in the tails or near structural boundaries such as $x_{\min}$.

Third, because parameters are estimated from the same data used to assess fit,
classical goodness-of-fit $p$-values are not valid.
We therefore employ a refit parametric bootstrap, {described in S2.3 below,} to properly calibrate
the null distribution of the test statistic.

The specific tools used in this layered framework are described below.




\paragraph{S3.2.1. Likelihood and BIC (relative fit).}
{For each candidate model, we computed the maximized log-likelihood
$\ell(\hat{\theta})$ and the Bayesian Information Criterion
\[
\mathrm{BIC} = -2\ell(\hat{\theta}) + k \log(n),
\]
where $k$ is the number of free parameters.
BIC penalizes additional parameters and favors parsimonious structure.

Across the families considered, models incorporating an explicit atom at
$x_{\min}$ consistently improved the likelihood relative to purely
continuous mixtures, reflecting the substantial empirical mass at the
measurement floor. Among the atom-based candidates, the
two-component continuous structure consisting of an
{E1-short} component (a short-duration density proportional to
$\exp(-rt)/t$ on $[x_{\min},\infty)$) and a {truncated normal (TN)}
long-duration component achieved lower BIC values than more flexible
three-component alternatives once the complexity penalty was taken into account.
This indicates that the Atom + E1-short + truncated-normal formulation provides
the most parsimonious representation of the data among the
interpretable candidate families considered.
Accordingly, subsequent analyses of goodness-of-fit diagnostics
and propensity estimation were based on this selected model.}

\paragraph{S3.2.2. Kolmogorov--Smirnov distance (absolute fit).}
To assess absolute agreement between the fitted model and the empirical
distribution, we computed the Kolmogorov--Smirnov (KS) distance
\[
D_{\mathrm{obs}} = \sup_t \big|F_n(t) - F_{\hat{\theta}}(t)\big|,
\]
{where $n$ is the number of observations, $F_n$ is the empirical CDF, and
$F_{\hat{\theta}}$ is the fitted model CDF (including the atom at $x_{\min}$).
The KS distance summarizes the largest vertical deviation between the two CDFs
and is sensitive to mismatches anywhere on the support, including near the boundary
and in the tail. We use $D_{\mathrm{obs}}$ as a global diagnostic of distributional
adequacy for the pooled dataset and for each REM--pre stratum.}

The KS distance summarizes the largest vertical deviation between the two CDFs,
and is sensitive to mismatches anywhere on the support, including near the boundary
and in the tail. We use $D_{\mathrm{obs}}$ as a global diagnostic of distributional adequacy for the pooled dataset and for each REM--pre stratum.

\paragraph{S3.2.3. Refit parametric bootstrap (calibrated $p$-values).}
When parameters are estimated from the same data used to compute the KS statistic,
the classical KS null distribution does not apply~\cite{stute1993bootstrap}.
To obtain calibrated $p$-values, we used a refit parametric bootstrap:

\begin{enumerate}[label=\arabic*., leftmargin=*]
\item Simulate a synthetic dataset of size $n$ from the fitted model $F_{\hat{\theta}}$
(including the atom at $x_{\min}$),

\item Refit the \emph{same} model to the synthetic data using the identical estimation
procedure (multi-start { expectation--maximization (EM) with the same constraints)}, obtaining $\hat{\theta}^{(b)}$. {Let $F_n^{(b)}$ denote the empirical CDF of the
$b$th synthetic sample.}
\item Compute the bootstrap KS statistic
\[
D^{(b)}=\sup_t \big|F_n^{(b)}(t) - F_{\hat{\theta}^{(b)}}(t)\big|,
\]
\item {Repeat steps 1--3 independently for $b=1,\dots,B$, and estimate}
\[
p = \frac{1 + \sum_{b=1}^B \mathbf{1}\{D^{(b)} \ge D_{\mathrm{obs}}\}}{B+1}.
\]
\end{enumerate}

This procedure properly accounts for parameter-estimation variability {because the model is re-estimated for each synthetic dataset before the KS statistic is recomputed. Consequently, the bootstrap null distribution reflects both sampling variability and the additional uncertainty introduced by parameter fitting}.
Goodness-of-fit results (including $D_{\mathrm{obs}}$ and refit-bootstrap $p$-values for the pooled sample and each REM--pre bin) are reported in \ref{tab:atom_fit_results}.

\subsection*{S3.3. Why the Atom + E1 + Truncated-Normal model was selected}

{Because the data are only observed for $t \ge x_{\min}$, there is no empirical information about behavior below the measurement floor. In particular, although the empirical CDF has a positive jump at $x_{\min}$, any hypothetical continuous behavior for $t<x_{\min}$ is not identifiable from the data and therefore should not be treated as part of the fitted continuous regime.}

Although LN+LN and certain three-component mixtures achieved lower BIC
than simple two-component EXP+LN or WEI+LN, the final Atom + E1 + TN model
was selected based on two overarching considerations:

\begin{enumerate}[label=(\roman*), leftmargin=*]
\item \textbf{Boundary-aware, interpretable decomposition.}
{The empirical distribution exhibits a pronounced lower-bound spike at the measurement floor $x_{\min}$ together with an approximately $1/t$-like decay in the density over short durations.
Purely continuous mixture models on $(0,\infty)$ do not accommodate this boundary structure well: a $1/t$-type behavior cannot extend to $0$ in a normalizable continuous density, whereas the observed data are naturally bounded below by $x_{\min}=0.5$ minutes. By truncating the continuous model to $[x_{\min},\infty)$ and assigning an explicit atom at $x_{\min}$, the model separates measurement-floor effects from the continuous short-duration regime. Conditional on $T>x_{\min}$, the E1-short component captures the observed near-boundary decay, while the truncated-normal component provides an interpretable representation of sustained NREM episodes.}

\item \textbf{Stability for downstream propensity analysis with minimal complexity.}
Propensity is highly sensitive to model behavior near
$x_{\min}$. Separating the atom prevents artificial lower-bound spikes in propensity curves.
At the same time, compared with more flexible three-component mixtures, the Atom + two
continuous components model retains interpretability while avoiding unnecessary complexity
that can reduce parameter stability and complicate propensity estimation.
\end{enumerate}

Thus, the final model balances boundary fidelity, interpretability, and stable downstream
propensity estimation.








\subsection*{S3.4. Final model specification}


\paragraph{Atom at the measurement floor.}
We explicitly model the large spike observed at $x_{\min}$ by assigning
a point mass (atom) at the floor:
\[
\Pr(T = x_{\min}) = a.
\]
Here $a\in(0,1)$ represents the fraction of cycles recorded exactly at
the floor value. Treating this spike as a discrete component prevents
the continuous densities from being distorted to accommodate an
artifact of discretization/measurement resolution.

\paragraph{Continuous mixture on $[x_{\min},\infty)$.}
Conditional on exceeding the floor ($T>x_{\min}$), durations are modeled
by a two-component continuous mixture:
\[
f_c(t) = w f_{\mathrm{E1}}(t;r)
+ (1-w) f_{\mathrm{TN}}(t;\mu,\sigma),
\qquad t \ge x_{\min}.
\]
The mixing weight $w\in(0,1)$ captures the probability that a
continuous duration belongs to the short-duration regime, while
$1-w$ corresponds to the long-duration regime.

\paragraph{Short component (E1-short).}
The short-duration component is designed to capture the high density
near $x_{\min}$ together with {an initially slowly decaying tail.}
It is defined on
$[x_{\min},\infty)$ by a normalized $\exp(-rt)/t$ form. The
normalization constant uses the exponential integral function
\[
E_1(z) \;=\; \int_{z}^{\infty}\frac{e^{-u}}{u}\,du,
\qquad z>0.
\]
With this choice, the
short-component PDF is
\[
f_{\mathrm{E1}}(t;r)
=
\frac{e^{-rt}}{t\,E_1(r x_{\min})},
\qquad t\ge x_{\min},
\]
which is properly normalized on $[x_{\min},\infty)$:
\[
\int_{x_{\min}}^{\infty} f_{\mathrm{E1}}(t;r)\,dt = 1.
\]
This component behaves like $1/t$ near the lower bound (supporting a
large near-$x_{\min}$ density) while retaining an exponential tail
$e^{-rt}$ that prevents unrealistically heavy long-duration mass.

\paragraph{Long component (truncated normal).}
Long durations are modeled by a normal distribution with mean $\mu$ and
standard deviation $\sigma$, truncated to the physically admissible
range $[x_{\min},\infty)$. Its pdf is
\[
f_{\mathrm{TN}}(t;\mu,\sigma)
=
\frac{\phi\!\left((t-\mu)/\sigma\right)}
{\sigma\Big(1-\Phi\!\left((x_{\min}-\mu)/\sigma\right)\Big)},
\qquad t\ge x_{\min},
\]
where $\phi$ and $\Phi$ are the standard normal pdf and CDF,
respectively. Truncation ensures the model assigns no probability mass
below the measurement floor, while $(\mu,\sigma)$ provide an interpretable
summary of the center and spread of sustained NREM episodes.

\subsection*{S3.5. Parameter estimation}

The atom mass $a$ is estimated directly as the empirical
proportion of observations equal to $x_{\min}$.
Continuous parameters $(w,r,\mu,\sigma)$ are estimated using
multi-start Expectation–Maximization (EM) applied to the subset
$\{x_i > x_{\min}\}$.
Multiple random initializations are used to reduce sensitivity
to local optima, and the solution with highest log-likelihood
is retained.

Given current parameters, posterior responsibilities for the
short component are computed as
\[
\gamma_i
=
\frac{w f_{\mathrm{E1}}(x_i;r)}
{w f_{\mathrm{E1}}(x_i;r)
+
(1-w)f_{\mathrm{TN}}(x_i;\mu,\sigma)}.
\]

Parameter updates are obtained via weighted maximum likelihood.
Because closed-form updates are not available for $r$,
numerical optimization is performed in the M-step.
To preserve identifiability and prevent degeneracy,
bounds are imposed on $r$, $\sigma$, and $\mu$,
and mixture weights are constrained to remain strictly between
0 and 1.
Convergence is declared when the relative change in
log-likelihood falls below a predefined tolerance.

\subsection*{S3.6. Goodness-of-fit results}
\label{supp:humanGOF}
{For each pooled and REM--pre stratified fit, we report the estimated atom mass $\hat{a}$ at the measurement floor $x_{\min}$, the continuous mixture weight $\hat{w}$ corresponding to the short-duration component, and the component parameters $(\hat{r},\hat{\mu},\hat{\sigma})$ governing the E1-short and truncated-normal long regimes. We additionally report the maximized log-likelihood and the Bayesian Information Criterion (BIC) for model comparison, together with the observed discrete Kolmogorov--Smirnov statistic and its calibrated refit-bootstrap $p$-value. These quantities jointly summarize relative model performance, absolute goodness-of-fit, and the structural decomposition of short- and long-duration NREM dynamics across pooled and REM--pre stratified analyses.

Because the observed NREMs durations $|N|$ are recorded exactly on a 0.5-min grid, goodness-of-fit was assessed using the corresponding \emph{discrete} model rather than by comparing the empirical sample directly to the latent continuous distribution. To account for possible ambiguity in how a latent continuous duration is represented on the observation grid, we evaluated three mapping rules: floor, ceil, and nearest. For the pooled human data, goodness-of-fit was computed under each of these mappings, and the rule used for the REM--pre stratified analyses was selected according to the largest refit-bootstrap KS $p$-value in the pooled analysis. Under this comparison, the \emph{ceil} mapping yielded the strongest agreement with the empirical grid-valued distribution and was therefore used for the REM--pre stratified goodness-of-fit calculations.

For a fixed mapping rule, the fitted atom + E1-short + truncated-normal model induces a probability mass function on the 0.5-min grid, from which a model-based discrete cumulative distribution function is obtained. The reported KS statistic is the maximum absolute difference between the empirical grid-based cumulative distribution function and this fitted discrete cumulative distribution function. The associated $p$-value was obtained by refit parametric bootstrap: synthetic datasets of the same size were generated from the fitted model, mapped to the grid using the same observation rule, refit by the same estimation procedure, and then re-evaluated using the same discrete KS statistic. This calibration accounts for parameter-estimation uncertainty and for the discretized nature of the observed data.

For the pooled human data, the observed discrete KS statistic under the selected \emph{ceil} mapping was $D_{\mathrm{obs}}=0.0168$, with refit-bootstrap $p=0.20$, indicating that the fitted model provides an adequate absolute description of the grid-valued empirical distribution. The same conclusion held across REM--pre strata. Specifically, for Bin~1 ($0.5 \le \mathrm{REMpre}<5$ min), $D_{\mathrm{obs}}=0.0237$ with $p=0.215$; for Bin~2 ($5 \le \mathrm{REMpre}<12$ min), $D_{\mathrm{obs}}=0.0194$ with $p=0.500$; and for Bin~3 ($12 \le \mathrm{REMpre}<71.5$ min), $D_{\mathrm{obs}}=0.0262$ with $p=0.115$. Thus, the observed KS discrepancies were typical under the fitted model once refitting uncertainty was taken into account, and the Atom + E1-short + truncated-normal model was not rejected as an absolute distributional description of the human $|N|$ data.

\begin{table}[H]
\centering
\caption{Goodness-of-fit summary for the Atom + E1-short + Truncated-Normal model. The reported KS statistic is the discrete Kolmogorov--Smirnov distance computed on the 0.5-min observation grid under the selected discretization rule (\emph{ceil}). All $p$-values were obtained from refit parametric bootstrap.}
\small
\begin{tabular}{lccccccccc}
\hline
\textbf{Fit} & $\hat a$ & $\hat w_{\text{short}}$ & $\hat r$ & $\hat \mu$ & $\hat \sigma$ & BIC & $D_{\mathrm{obs}}$ & $p$ \\
\hline
Lumped 
& 0.2922 
& 0.7481 
& 0.1081 
& 65.5819 
& 16.6177 
& 11046.05 
& 0.0168 
& 0.200 \\

Bin 1 $[0.5,5)$ 
& 0.3541 
& 0.8753 
& 0.1329 
& 58.8130 
& 16.6844 
& 3938.06 
& 0.0237 
& 0.215 \\

Bin 2 $[5,12)$ 
& 0.3056 
& 0.7733 
& 0.1126 
& 64.0311 
& 16.4390 
& 3140.80 
& 0.0194 
& 0.500 \\

Bin 3 $[12,71.5)$ 
& 0.1470 
& 0.5231 
& 0.0630 
& 69.4299 
& 16.2589 
& 3812.39 
& 0.0262 
& 0.115 \\
\hline
\end{tabular}
\normalsize
\label{tab:atom_fit_results}
\end{table}

}

\subsection*{S3.7. Propensity estimation}

We define discrete-time propensity over a window $\Delta$ as
\[
p_\Delta(t)
=
\frac{F_c(t+\Delta)-F_c(t)}{1-F_c(t)},
\]
where $F_c$ denotes the CDF of the continuous component only.
Throughout, $\Delta = 0.5$ minutes (30 seconds).

\paragraph{Why the continuous CDF is used.}

The full fitted CDF has the form
\[
F(t) =
\begin{cases}
0, & t < x_{\min},\\
a + (1-a)F_c(t), & t \ge x_{\min},
\end{cases}
\]
where $a$ is the atom mass at $x_{\min}$.
{For the propensity analysis, however, we are interested in termination dynamics conditional on the episode having already exceeded the measurement floor.
Accordingly, for times $t>x_{\min}$, the relevant distributional behavior is
described by the continuous component $F_c$, not by the atom at $x_{\min}$.}
If the full CDF $F$ were used directly in the propensity definition,
then near $t = x_{\min}$ the numerator would contain the discrete jump
$a$, producing
\[
F(t+\Delta)-F(t)
\approx a + \text{(continuous contribution)}.
\]
Because the survival term in the denominator is also altered,
this discrete jump can generate an artificial spike in
\[
p_\Delta(t)
=
\frac{F(t+\Delta)-F(t)}{1-F(t)},
\]
immediately above $x_{\min}$.
This spike does not represent genuine termination dynamics of
continuous NREM durations; rather, it reflects discretization
and measurement-floor effects.
Including the atom in the propensity calculation therefore
conflates two distinct mechanisms: (1) the structural measurement-floor mass at $x_{\min}$, and
(2) the intrinsic termination dynamics of continuous NREM episodes.

To isolate the latter, we compute propensity using $F_c$,
the CDF of the continuous mixture only.
This ensures that propensity reflects
\[
\Pr(T \in [t,t+\Delta) \mid T \ge t,\ T>x_{\min}),
\]
\emph{i.e.}, termination dynamics conditional on having already
exceeded the measurement floor.
Note this modification affects only the propensity calculation, not the fitted model itself.
Even when using $F_c$, numerical instability can arise exactly at
$t=x_{\min}$ because of truncation.
We therefore evaluate propensity beginning at
\[
t_{\mathrm{start}} = x_{\min} + 0.05,
\]
where $0.05$ minutes corresponds to 3 seconds.
This small offset removes residual boundary artifacts
without altering the substantive shape of the curve.

\section{Hypothesis tests for normalized-time REM distributions and onset-decile REM bout duration (Figures~\ref{fig:percent_time_bins}--~\ref{fig:seq_single})}\label{sec: supp_fig_9}

To complement the descriptive bar plots, we carried out subject-level permutation-based hypothesis tests designed to assess whether the observed decile structure reflects a reproducible pattern across subjects rather than only pooled event counts. Because these analyses are based on repeated measurements within subject, we avoided tests that would treat all REM bouts as independent observations.

\subsection*{Edge-versus-middle enrichment of REM onset fractions}

Rather than fitting a linear trend across normalized-time deciles, we tested whether REMS onset fractions near the beginning and end of the sleep episode differed from the corresponding fractions in the middle of the sleep episode. For each subject \(i\), let $x_{i,1}$
denote the fraction of that subject's nightly REMS onsets falling in the first decile (0--10\% of normalized sleep time), let $x_{i,10}$
denote the fraction falling in the last decile (90--100\%), and define the subject-specific middle-decile mean by
\[
\bar{x}_{i,\mathrm{mid}}=\frac{1}{8}\sum_{d=2}^{9} x_{i,d}.
\]
We then formed three paired contrasts:
\[
d_i^{(1)} = x_{i,1} - \bar{x}_{i,\mathrm{mid}},
\]
\[
d_i^{(10)} = x_{i,10} - \bar{x}_{i,\mathrm{mid}},
\]
and
\[
d_i^{(\mathrm{edge})} = \frac{x_{i,1}+x_{i,10}}{2} - \bar{x}_{i,\mathrm{mid}}.
\]

For each contrast, the null hypothesis was that the mean paired difference across subjects was zero,
\[
H_0:\ \mathbb{E}[d_i]=0,
\]
against the two-sided alternative
\[
H_A:\ \mathbb{E}[d_i]\neq 0.
\]

To evaluate these hypotheses without imposing a Gaussian assumption on subject-level fractions, we used a two-sided paired sign-flip permutation test. For a given set of subject-level paired differences \(d_1,\dots,d_n\), the observed test statistic was the sample mean
\[
T_{\mathrm{obs}} = \frac{1}{n}\sum_{i=1}^n d_i.
\]
Under the null hypothesis of no systematic edge-versus-middle effect, the sign of each paired difference is exchangeable. We therefore generated a null distribution by repeatedly multiplying each \(d_i\) by an independent random sign \(s_i\in\{-1,+1\}\) with equal probability and recomputing
\[
T^{(b)} = \frac{1}{n}\sum_{i=1}^n s_i d_i,
\qquad b=1,\dots,B.
\]
Using \(B=20{,}000\) random sign-flip resamples, the two-sided permutation \(p\)-value was estimated as
\[
p = \frac{1}{B}\sum_{b=1}^B \mathbf{1}\!\left(|T^{(b)}|\ge |T_{\mathrm{obs}}|\right).
\]
This procedure was applied separately to the first-decile contrast, the last-decile contrast, and the averaged-edge contrast.

These tests assess whether the subject-level REM onset fraction at the beginning of the night, at the end of the night, or averaged across both the first and last decile differs from the corresponding subject-level average fraction across the middle 10--90\% of normalized sleep time. Because the analysis is performed at the subject level, the inference reflects between-subject consistency of REMS enrichment  at the beginning and end of the sleep period rather than only pooled event counts.

\subsection*{Association between REM bout duration and onset decile}

We next tested whether REM bout duration varies across normalized sleep-time onset deciles. Although the corresponding figure displays pooled bout-level means, direct inference at the bout level would treat multiple bouts from the same subject as independent observations. To avoid this pseudo-replication, we used a subject-level within-subject permutation framework.

For each subject \(i\) and onset decile \(d\), we computed the mean REM bout duration among that subject's bouts assigned to decile \(d\):
\[
m_{i,d} = \frac{1}{n_{i,d}}\sum_{c \in \mathcal{B}_{i,d}} Y_c,
\]
where \(\mathcal{B}_{i,d}\) denotes the set of bouts from subject \(i\) in decile \(d\), \(n_{i,d}=|\mathcal{B}_{i,d}|\), and \(Y_c\) is the REM bout duration of bout \(c\). If a subject had no bouts in a given decile, the corresponding entry was treated as missing. This produced a subject-by-decile matrix
\[
M = (m_{i,d}),
\]
with one row per subject and one column per onset decile.

We tested the global null hypothesis that REM bout duration is unrelated to onset decile:
\[
H_0:\ \text{within each subject, onset-decile labels carry no information about REM bout duration,}
\]
against the alternative
\[
H_A:\ \text{REM bout duration depends on onset decile.}
\]
Equivalently, under \(H_0\), onset-decile labels are exchangeable within subject.

To quantify decile-dependent structure while controlling for subject-specific baseline differences in REM bout duration, we first centered each subject's row by subtracting that subject's mean across observed deciles:
\[
\tilde{m}_{i,d} = m_{i,d} - \bar{m}_i,
\qquad
\bar{m}_i = \frac{1}{|\mathcal{D}_i|}\sum_{d\in\mathcal{D}_i} m_{i,d},
\]
where \(\mathcal{D}_i\) is the set of onset deciles observed for subject \(i\). We then computed the decile-specific centered mean across subjects,
\[
\tilde{\mu}_d = \frac{1}{N_d}\sum_{i:\, d\in\mathcal{D}_i} \tilde{m}_{i,d},
\]
where \(N_d\) is the number of subjects contributing data to decile \(d\). The observed test statistic was the summed squared decile effect,
\[
T_{\mathrm{obs}} = \sum_{d=1}^{10} \tilde{\mu}_d^2.
\]
Large values of \(T_{\mathrm{obs}}\) indicate systematic variation in subject-centered REM bout duration across onset deciles.

To generate the null distribution, we permuted onset-decile labels \emph{within each subject}; that is, for each subject, we randomly reassigned that subject's observed decile labels across that subject's bouts while leaving the REM bout durations unchanged. This preserves the number of bouts contributed by each subject, the subject-specific distribution of REM bout durations, and the marginal number of observations per subject, while destroying any systematic association between REM bout duration and onset decile within subject.

For each permutation \(b=1,\dots,B\), we recomputed the subject-level decile means, the centered decile effects, and the corresponding test statistic \(T^{(b)}\). Using \(B=20{,}000\) within-subject permutations, the permutation \(p\)-value was estimated as
\[
p = \frac{1}{B}\sum_{b=1}^{B}\mathbf{1}\!\left(T^{(b)} \ge T_{\mathrm{obs}}\right).
\]

This test evaluates whether the decile at which a REM bout begins is associated with REM bout duration, after accounting for subject-specific baseline differences and without assuming a linear trend across the night. A significant result indicates that REM bout duration varies across onset deciles, but does not by itself imply a monotone increase or decrease.

\end{document}